\documentclass{emulateapj}
\usepackage{amssymb}
\usepackage{graphicx}
\usepackage{epstopdf}
\usepackage{url}

\newcommand{\msun}{\ensuremath{\rm M_\odot}}

\newcommand{\msunyr}{\ensuremath{\rm M_{\odot}\;{\rm yr}^{-1}}}
\newcommand{\Ha}{\ensuremath{\rm H\alpha}}
\newcommand{\Hb}{\ensuremath{\rm H\beta}}
\newcommand{\lya}{\ensuremath{\rm Ly\alpha}}

\newcommand{\kms}{km~s\ensuremath{^{-1}}}

\newcommand{\ztwo}{\ensuremath{z\sim2}}

\newcommand{\OII}{[\ion{O}{2}]}
\newcommand{\OIII}{[\ion{O}{3}]}

\newcommand{\FeII}{\ion{Fe}{2}}
\newcommand{\MgII}{\ion{Mg}{2}}
\newcommand{\MgI}{\ion{Mg}{1}}
\newcommand{\HII}{\ion{H}{2}}
\newcommand{\SiII}{\ion{Si}{2}}
\newcommand{\CII}{\ion{C}{2}}
\newcommand{\CIIsf}{\ion{C}{2}]}

\newcommand{\Wo}{$W_0$}

\begin{document}

\title{Galactic Outflows in Absorption and Emission:\\ Near-UV Spectroscopy of Galaxies at $1<z<2$\altaffilmark{*}}
\author{{\sc Dawn K. Erb}\altaffilmark{1}, {\sc Anna M. Quider}\altaffilmark{2},  {\sc Alaina L. Henry}\altaffilmark{3}, and {\sc  Crystal L. Martin}\altaffilmark{3}}

\slugcomment{Accepted for publication in \apj}

\shorttitle{GALACTIC OUTFLOWS IN ABSORPTION AND EMISSION}
\shortauthors{ERB, QUIDER, HENRY AND MARTIN}

\altaffiltext{*}{Based on data obtained at the W. M. Keck Observatory, which is operated as a scientific partnership among the California Institute of Technology, the University of California, and the National Aeronautics and Space Administration, and was made possible by the generous financial support of the W. M. Keck Foundation.}

\altaffiltext{1}{University of Wisconsin Milwaukee, Department of Physics, Milwaukee, WI 53211, USA; erbd@uwm.edu}
\altaffiltext{2}{Institute of Astronomy, Madingley Road, Cambridge CB3 0HA, UK; aquider@gmail.com}
\altaffiltext{3}{University of California Santa Barbara, Department of Physics, Santa Barbara, CA 93106, USA; ahenry@physics.ucsb.edu, cmartin@physics.ucsb.edu}

\begin{abstract}
We study large-scale outflows in a sample of 96 star-forming galaxies at $1\lesssim z \lesssim 2$, using near-UV spectroscopy of \FeII\ and \MgII\ absorption and emission.  The average blueshift of the \FeII\ interstellar absorption lines with respect to the systemic velocity is $-85\pm10$ \kms\ at $z\sim1.5$, with standard deviation 87 \kms; this is a decrease of a factor of two from the average blueshift measured for far-UV interstellar absorption lines in similarly selected galaxies at \ztwo.  The profiles of the \MgII\ $\lambda\lambda$2796, 2803 lines show much more variety than the \FeII\ profiles, which are always seen in absorption; \MgII\ ranges from strong emission to pure absorption, with emission more common in galaxies with blue UV slopes and at lower stellar masses. Outflow velocities, as traced by the centroids and maximum extent of the absorption lines, increase with increasing stellar mass with 2--3$\sigma$ significance, in agreement with previous results. We study fine structure emission from \FeII*, finding several lines of evidence in support of the model in which this emission is generated by the re-emission of continuum photons absorbed in the \FeII\ resonance transitions in outflowing gas.  In contrast, photoionization models indicate that \MgII\ emission arises from the resonant scattering of photons produced in \HII\ regions, accounting for the differing profiles of the \MgII\ and \FeII\ lines.  A comparison of the strengths of the \FeII\ absorption and \FeII* emission lines indicates that massive galaxies have more extended outflows and/or greater extinction, while two-dimensional composite spectra indicate that emission from the outflow is stronger at a radius of $\sim$10 kpc in high mass galaxies than in low mass galaxies.
\end{abstract}


\section{Introduction}
\label{sec:intro}
The evolution of galaxies cannot be understood without the simultaneous consideration of the surrounding gas.  Galaxies drive powerful, galactic-scale outflows, and must also accrete gas from the intergalactic medium (IGM).  Thus the process of galaxy evolution reflects the cycling of baryons in and out of galaxies.

The importance of galactic outflows is supported by a wide range of models and observations.  An incomplete list of phenomena that may be explained by galactic outflows includes the shape of the mass-metallicity relation and the metal enrichment of the IGM (e.g.\ \citealt{garn02,thk+04,od06,esp+06,d07,fd07,bgb+07,mcm+09,ps11,mwn+11}), the quenching of star formation in massive galaxies (e.g.\ \citealt{tmd07,hckh08,gdof11}), and the mismatch between the galaxy luminosity function and the mass function of dark matter halos (e.g.\ \citealt{bbf+03,kkd+09,odk+10}).  Powerful galactic outflows must also modulate the inflow of gas into galaxies, constraining the assembly of the baryonic component and regulating star formation (e.g.\ \citealt{dof11,fkm11,vsat12}).

In the broadest terms, galactic outflows are powered by energy and momentum injected into the interstellar medium (ISM) of galaxies by star formation and/or active galactic nuclei (AGN).  Hot ($\sim10^7$ K), metal-enriched supernova ejecta may be expelled from galaxies, entraining significant amounts of the cool ($\sim10^4$ K) ISM in the process.   Even before massive stars explode as supernovae, radiation pressure may accelerate the cool gas to large velocities \citep{mmt11}.  The detailed physics of the outflows remains controversial, however, and observational constraints on the velocity of outflows as a function of radius, the mass outflow rate, the spatial extent of outflows, and the scaling of these quantities with galaxy mass and star formation rate are needed.

\begin{figure*}[htbp]
\plotone{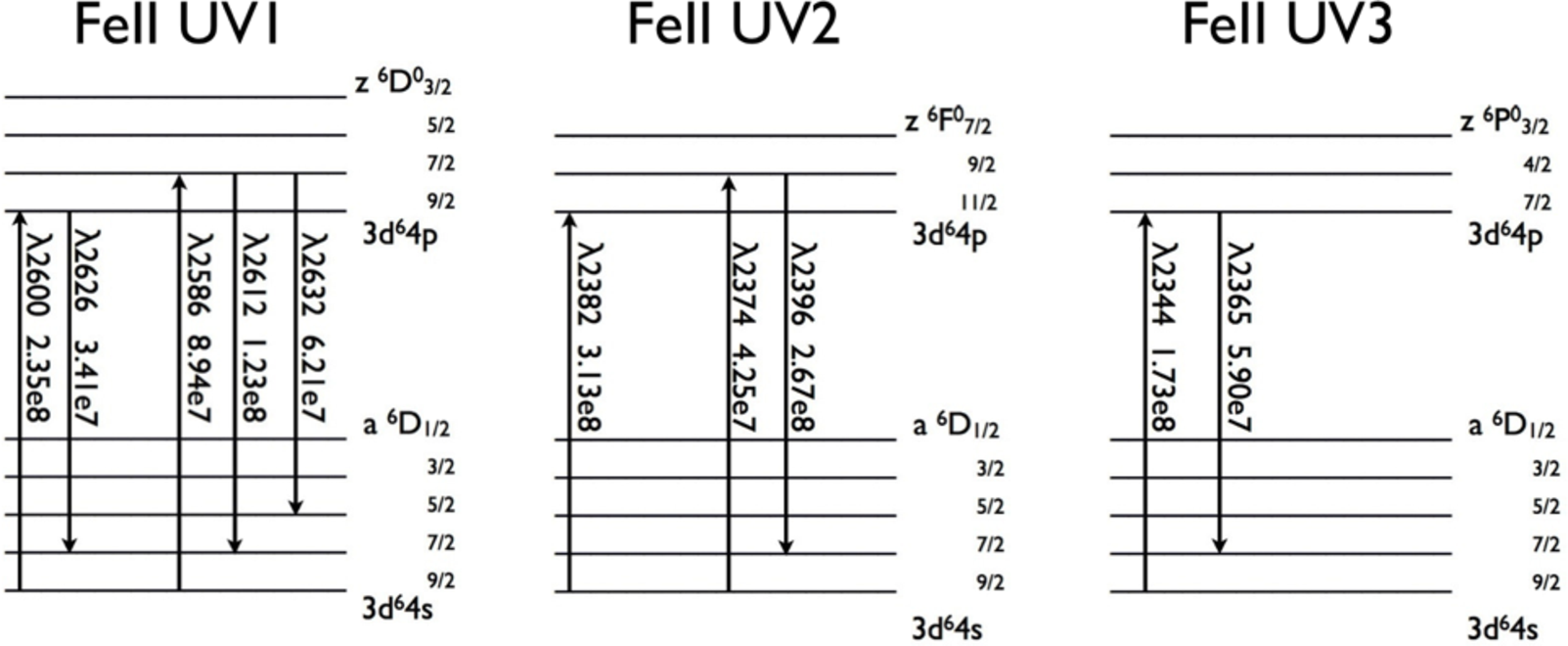}
\caption{The energy level diagrams for selected transitions of the \FeII\ atom (based on Figure 7 of \citealt{hmt+99} and Figure 1 of \citealt{pkr11}). Each transition is labeled with its rest wavelength (\AA) and Einstein A-coefficient (s$^{-1}$). Upward arrows indicate resonance absorption (absorption from the ground state). The fine structure splitting of the energy levels allows for fine structure emission. The downward arrows show the \FeII* (fine structure) emission that is energetically tied to the resonance absorption transitions. Only \FeII* transitions linked to the resonant absorption transitions are shown. }
\label{fig:energylevels}
\end{figure*}

Galactic winds are most commonly traced via the blueshifts of interstellar absorption lines from cool, outflowing gas, seen against the backlight of the stellar continuum.  In the local universe, outflows are found in starburst galaxies, from dwarf starbursts to ULIRGs (e.g.\ \citealt{ham90,ss02,vcb05,htb+07,mb09,smpa12}).  Both observations and models suggest a critical star formation rate density in order to drive outflows \citep{h02,mmt11,ksm+12}, and outflow velocity is observed to increase with galaxy mass and star formation rate \citep{m05,rvs05}.

At higher redshifts, $1.5 \lesssim z \lesssim 4$, outflows are ubiquitous, with most galaxies observed showing absorption lines blueshifted by $\sim100$-300 \kms\ (e.g.\ \citealt{ssp+03, wcp+09, ses+10, jse12}).  Outflows in high redshift galaxies are also traced by \lya, which is seen in redshifted emission as well as blueshifted absorption; the emission is interpreted as resonant scattering of \lya\ photons from receding gas on the far side of the galaxy.  Because the high redshift studies have been restricted to a smaller range in mass and star formation rate than the local samples (which extend to low mass dwarfs), trends between outflow properties and galaxy properties have been more difficult to determine; however, recent results at $z\sim1.4$ suggest similar scalings with galaxy mass and SFR \citep{wcp+09}.

A major difficulty in the study of outflows is the measurement of their spatial extent, since absorption line studies of galaxies offer no information on the location of the absorbing gas.  Spatial information can be obtained by measuring absorption in the spectra of background objects (QSOs or galaxies) with lines of sight passing near galaxies. \citet{ses+10} used composite spectra of pairs of foreground and background galaxies, stacked as a function of impact parameter, to show that significant metal line absorption is seen around galaxies to distances of $\sim 60$ kpc, while \lya\ absorption extends to $\sim 250$ kpc.  Searches for star-forming counterparts to \MgII\ absorption systems also suggest that galaxies drive outflows to large distances \citep{bmp+07,mwn+11}, while the observation that strong \MgII\ absorption systems are preferentially found along the minor axis of disk galaxies is further evidence that these systems are associated with outflows \citep{kcn12,bhv+12}. On the other hand, not all absorption systems have detectable counterparts \citep{bmp+12} and the absorption has also been interpreted as being due to infalling gas \citep{cwt+10}, a scenario potentially supported by the fact that an increase in \MgII\ absorption is also seen along the major axes of disk galaxies  \citep{kcn12,bhv+12}.

The extent of outflowing gas can also be measured if galactic winds can be observed in emission.  Extended \Ha\ and X-ray emission is observed in local starbursts \citep{lhw99}, but such emission becomes undetectable at higher redshifts.  At $z=0.69$, \citet{rpm+11} detect \MgII\ emission at a radius of $\sim 7$ kpc from a starburst galaxy, while \citet{sbs+11} see diffuse \lya\ halos extending to $\sim 80$ kpc around star-forming galaxies at $z\sim2$--3, using stacked narrow-band images.  Both the \MgII\ and the \lya\ photons are interpreted as arising from resonant scattering in outflowing gas.

In this study we focus on \MgII\ and \FeII\ absorption and emission in galaxies at $1 \lesssim z \lesssim 2$.  The \MgII\ $\lambda\lambda$ 2796, 2803 doublet is a particularly useful diagnostic because it can be observed from the ground over a wide range in redshift, from $z\sim 0.2$ to \ztwo, thus in principle offering direct comparisons of outflow properties over $\sim8$ Gyr of cosmic time.  In practice, samples across this redshift range have so far differed significantly in size, selection technique, and type of galaxy observed \citep{tmd07,wcp+09,mb09,rwk+10,rpm+11,cwh+11,ksm+12,msc+12,dmt+12}.  

The study of \MgII\ and \FeII\ absorption and emission has also benefited from recent radiative transfer modeling of these transitions.  \citet{pkr11} use a variety of different galactic outflow models to predict the strengths of these lines, emphasizing the importance of photon scattering and the effect of re-emitted photons on the line profiles. Photons absorbed in resonance lines such as \MgII\ are re-emitted near the systemic velocity, creating underlying emission which fills in the absorption and may affect inferences about the velocity and covering fraction of outflowing gas.  Because most of the \FeII\ resonance lines are coupled to fine structure transitions, as shown in the energy level diagram in Figure \ref{fig:energylevels}, the photons absorbed in the \FeII\ transitions may instead be re-emitted in the \FeII* fine structure lines.  This paper focuses on the effect of photon scattering on measurements of absorption line velocity centroids and equivalent widths, and on \FeII* and \MgII\ emission and absorption as tracers of outflowing gas.

The paper is organized as follows.  We describe our sample and the acquisition and analysis of the data in Section \ref{sec:obs}, the determination of systemic redshifts in Section \ref{sec:zsys}, and the determination of stellar population properties in Section \ref{sec:sedfitting}.  Results from individual spectra are reported in Section \ref{sec:individual}: in Section \ref{sec:mg2uvcolor} we discuss \MgII\ emission, and Section \ref{sec:velw_indspec} compares the velocities and equivalent widths of the \FeII\ and \MgII\ transitions.  In Section \ref{sec:composites} we construct composite spectra based on stellar population properties (Section \ref{sec:compspecs}) and on the presence or absence of \MgII\ emission.  We discuss the velocity centroids and equivalent widths of the transitions in Sections \ref{sec:vel_compspec} and  \ref{sec:w_compspec} and the maximum observed outflow velocities in Section \ref{sec:maxvels}.  The remainder of the paper, Section \ref{sec:emission}, is devoted to the examination of \FeII* and \MgII\ emission.  In Section \ref{sec:summary} we summarize our results.  We use a cosmology with  $H_0=70\;{\rm km}\;{\rm s}^{-1}\;{\rm Mpc}^{-1}$, $\Omega_m=0.3$, and $\Omega_{\Lambda}=0.7$.

\section{Observations and Data Reduction}  
\label{sec:obs}
The galaxies observed in this study were drawn from the survey described by \citet{ssp+04}.  Targets were photometrically pre-selected using the rest-frame UV ``BM" and ``BX" color critieria; the BM criteria target galaxies in the redshift range $1.4 \lesssim z \lesssim 2$, and the BX criteria select galaxies primarily at $2 \lesssim z \lesssim 2.5$.  We observed galaxies in the Q1623 ($\alpha=16$:25:45, $\delta=+26$:47:23), Q1700 ($\alpha=17$:01:01, 	$\delta=+64$:11:58) and Q2343 ($\alpha=23$:46:05, $\delta=	+12$:49:12) survey fields. 

\begin{deluxetable*}{l l c c c c}
\tablewidth{0pt}
\tabletypesize{\footnotesize}
\tablecaption{Observations\label{tab:obs}}
\tablehead{
\colhead{Field} &
\colhead{Date} & 
\colhead{Exposure Time} & 
\colhead{$\lambda_{\rm central}$} & 
\colhead{Slit Width} &
\colhead{Blocking Filter} \\
\colhead{} &
\colhead{(UT)} &
\colhead{(s)} &
\colhead{(\AA)} &
\colhead{(arcsec)} &
\colhead{} 
}
\startdata
Q1623 & 02-04 Aug 2008 & 34,200 & 7500 & 1.0 & OG550 \\
Q1700 & 22-24 Aug 2009 & 27,000 & 7200 & 1.2 & GG495 \\
Q2343 & 02 Aug 2008; 21-24 Aug 2009 & 39,600 & 7200 & 1.2 & OG550 
\enddata
\end{deluxetable*}

Observations were conducted with the DEep Imaging Multi-Object Spectrograph (DEIMOS; \citealt{deimos}) on the Keck II telescope in August 2008 and August 2009.  DEIMOS is a medium resolution optical spectrometer with spectral coverage from $\sim$ 4100 $-$ 11,000 \AA; this  range includes rest-frame near-UV features from \FeII\,$\lambda\,2260$ to \OII $\lambda\lambda\,3727,3729$ in galaxies with $1 \leq z \leq 2$.  With the use of a slitmask, the 16.7\arcmin\ $\times$ 5.0\arcmin\ field of view allows the simultaneous observation of up to $\sim100$ galaxies. 

We observed one mask in each of the three fields, covering a total of 147 galaxies.  Galaxies were selected for the masks based on the following priorities: 1) known redshift (from the previous far-UV spectroscopy of \citealt{ssp+04}) in the range $1.4\lesssim z \lesssim 2$; 2) the BM (as opposed to BX) color criteria, which target the redshift range for which features of interest fall within the spectral co verage; and 3) brighter galaxies with ${\cal R} \lesssim 24$.  We used the 600 l/mm grating, which gives a dispersion of 0.65 \AA/pixel.  Exposure times ranged from 7.5 to 11 hrs, with average seeing FWHM~$\sim0.6$\arcsec. Details of the observations are given in Table \ref{tab:obs}.

Most of the data reduction was done with the DEEP2 data processing pipeline \citep{ncd+12,cnd+12},\footnote{\url{http://www2.keck.hawaii.edu/inst/deimos/pipeline.html}} which produces one-dimensional air wavelength-calibrated galaxy, inverse variance and sky spectra.  We then removed the instrumental signature from the data by putting the galaxy and error spectra on a relative flux scale with the IRAF software package, using standard stars observed in the same manner as the galaxies. 

We also removed the strong telluric absorption bands due to molecular oxygen at $\sim 6800$\,\AA\ and $\sim 7500$\,\AA.  This correction is usually done by scaling the strength of the telluric absorption seen in a standard star spectrum to the absorption strength seen in an object's spectrum. Because of the long integration times used in this study, however, the galaxies were observed over a range of airmasses, making a single standard star observation taken at a single air mass a poor match to the telluric feature strength in the individual galaxy spectra. A better measurement of the shape of the telluric features can be obtained from an average observed-frame galaxy spectrum for each field (because the galaxies lie at different redshifts, combining in the observed frame averages over the galaxy properties but gives a good measurement of the sky features).  We used these average spectra to create telluric feature templates for each field. Each average galaxy spectrum was normalized and smoothed, and then portions of the spectrum which were outside of the telluric bands were forced to have a normalized flux value of one, leaving the telluric bands as the only deviations from continuum. The resulting spectrum served as the template telluric spectrum for each field. This template was then divided into each of the galaxy and error spectra. 
 
Finally, the individual spectra were converted to a vacuum wavelength scale and rebinned by a factor of two to increase S/N; the 600 l/mm grating is DEIMOS' lowest resolution mode and so results in oversampled data.  The final pixel scale is 1.25 \AA\ per pixel in the observed frame, and the spectral resolution (measured from the widths of the night sky lines at $\sim 7500$ \AA) is $\sim 3.7$ \AA\ (FWHM) or $\sim 150$ \kms\ in the Q1623 field, which was observed with 1\arcsec\ slits, and $\sim 4.5$ \AA\ or $\sim 175$ \kms\ in the Q1700 and Q2343 fields, which were observed with 1.2\arcsec\ slits.

We measured redshifts $z>1$ for 96 of the 147 galaxies observed, either from \OII $\lambda\lambda$3727, 3729 emission or \FeII\ and \MgII\ absorption as discussed in Section \ref{sec:zsys} below (the remaining 51 galaxies were mostly among the fainter objects on the masks, and most had spectra too noisy to enable redshift determination; 4 galaxies had $z<1$).  The analysis presented here is based on this sample of 96 galaxies.  

After measuring redshifts, the 96 spectra were normalized using the procedure outlined by \citet{rpl+04}.  Normalization was done only in the rest-frame wavelength range 2200--3000 \AA; this range contains the \FeII\ and \MgII\ features of interest, and increasing sky noise at redder wavelengths makes normalization more difficult.  We shifted the spectra into the rest frame, identified regions of the spectrum which contained flux solely from the continuum, and fit a spline curve to both the mean wavelength and mean flux values within these defined continuum windows. The same windows were used for each galaxy, with adjustment on an individual basis to avoid the gap between detectors or noise spikes. Each galaxy spectrum was then divided by the spline fit to its continuum.  The uncertainty in the continuum fits is $\sim \pm 5\,\%$ for most galaxies.   These continuum normalized spectra were used for the remainder of the analysis.

We also make use of multi-wavelength imaging in order to model the spectral energy distributions of the galaxies.  The optical images, in the $U_n$, $G$ and ${\cal R}$ filters, are described by \citet{ssp+04},  the near-IR ($J$ and $K_s$) images by \citet{ess+06mass}, and the mid-IR {\it Spitzer} IRAC data by \citet{sse+05} and  \citet{rse+06,rps+12}.

\section{Determination of Systemic Redshift}
\label{sec:zsys}
The systemic redshift of a galaxy (i.e., the redshift of its stars) serves as the velocity zero-point for studying the kinematics of galactic and circumgalactic gas, and is used to align galaxies in the rest-frame for the construction of composite spectra, as described in Section \ref{sec:composites}.  A reliable determination of the systemic redshift is therefore required for this study.

The preferred method of redshift determination, measurement of the wavelengths of strong nebular emission lines from \HII\ regions, becomes increasingly difficult for galaxies with $z\gtrsim1.5$, as these lines redshift into the near-IR.  The centers of stellar photospheric absorption lines can also be used, but while these features are located at accessible wavelengths, they are broad and shallow and therefore hard to identify at the S/N of most high redshift galaxy spectra. If none of these features are available, the systemic redshift can be estimated from the wavelengths of the interstellar absorption lines.

In this study we use the first and third of the above methods.  For 51 of the 96 galaxies in the sample, strong \OII\ $\lambda\lambda$3727, 3729 emission is included at the red end of the spectral region covered, allowing a precise determination of the systemic redshift from a double Gaussian fit to the \OII\ doublet, which is usually resolved.  These 51 galaxies have a mean redshift $\langle z \rangle = 1.399 \pm 0.20$.

For the remaining 45 galaxies in the sample, we turn to the interstellar absorption lines.  It is well-documented that these features are blueshifted with respect to the systemic redshift and that, while the degree of blueshifting varies for individual galaxies, the assumption of an average velocity offset between the systemic and interstellar absorption redshifts allows the construction of composite spectra which are effectively at zero velocity \citep{ssp+03, ses+10}.  We follow the method of \citet{ses+10}, who use a sample of star-forming galaxies at $z \sim 2.2$  with both interstellar absorption lines and systemic redshifts from \Ha\ emission to find that, on average, the interstellar absorption lines are offset from the systemic redshift by $\Delta v = -166 \pm 130$ \kms.  

We use the subsample of galaxies with measurements of both \OII\ emission and \FeII\ $\lambda$2374 absorption to derive the average offset between nebular emission and interstellar absorption for our sample (as discussed in Sections \ref{sec:individual} and \ref{sec:composites}, this \FeII\ transition is expected to be least affected by the presence of emission, and therefore most representative of the kinematics of the outflow).  As described in more detail in Section \ref{sec:individual}, absorption line centroids are measured from a single Gaussian fit using the IDL routine {\it mpfit.pro}, which measures the center and equivalent width of each feature and their associated uncertainties.  Because \OII\ and this \FeII\ transition are at opposite ends of the accessible wavelength range, there are 20 galaxies whose spectra include significant detections of both features; the average uncertainty in the \FeII\ velocity measurements of this sample is 41 \kms.  The relationship between the \OII-defined systemic redshift and the redshift of the interstellar absorption is given by
\begin{equation}
z_{\rm{sys}} = z_{\rm{abs}} - \frac{\Delta v_{\rm abs}}{c} (1+z_{\rm abs}),
\label{eq:zsys}
\end{equation}
where $\Delta v_{\rm abs} = -85$ \kms\ is the average velocity offset (we compute an unweighted mean to avoid the possibility of biasing the result toward brighter galaxies, which may also be more massive). The standard deviation is 87 \kms, reflecting the significant scatter among individual measurements, while the error in the mean is 10 \kms.

For the higher S/N spectra with systemic redshifts from \OII\ emission, we also measure velocity offsets through a joint fit to the \FeII\ $\lambda$2344, 2374 and 2587 lines (for more details, see \citealt{msc+12}).  The mean absorption line velocity measured in this way is $\Delta v = -76\pm6$ \kms\ with a standard deviation of 80 \kms, in good agreement with the value of $-85$ \kms\ measured from \FeII\ $\lambda$2374 alone;  the joint and \FeII\ $\lambda$2374 velocities agree within 2$\sigma$ in all cases, and within 1$\sigma$ in half of the measured sample.

\begin{figure}[htbp]
\plotone{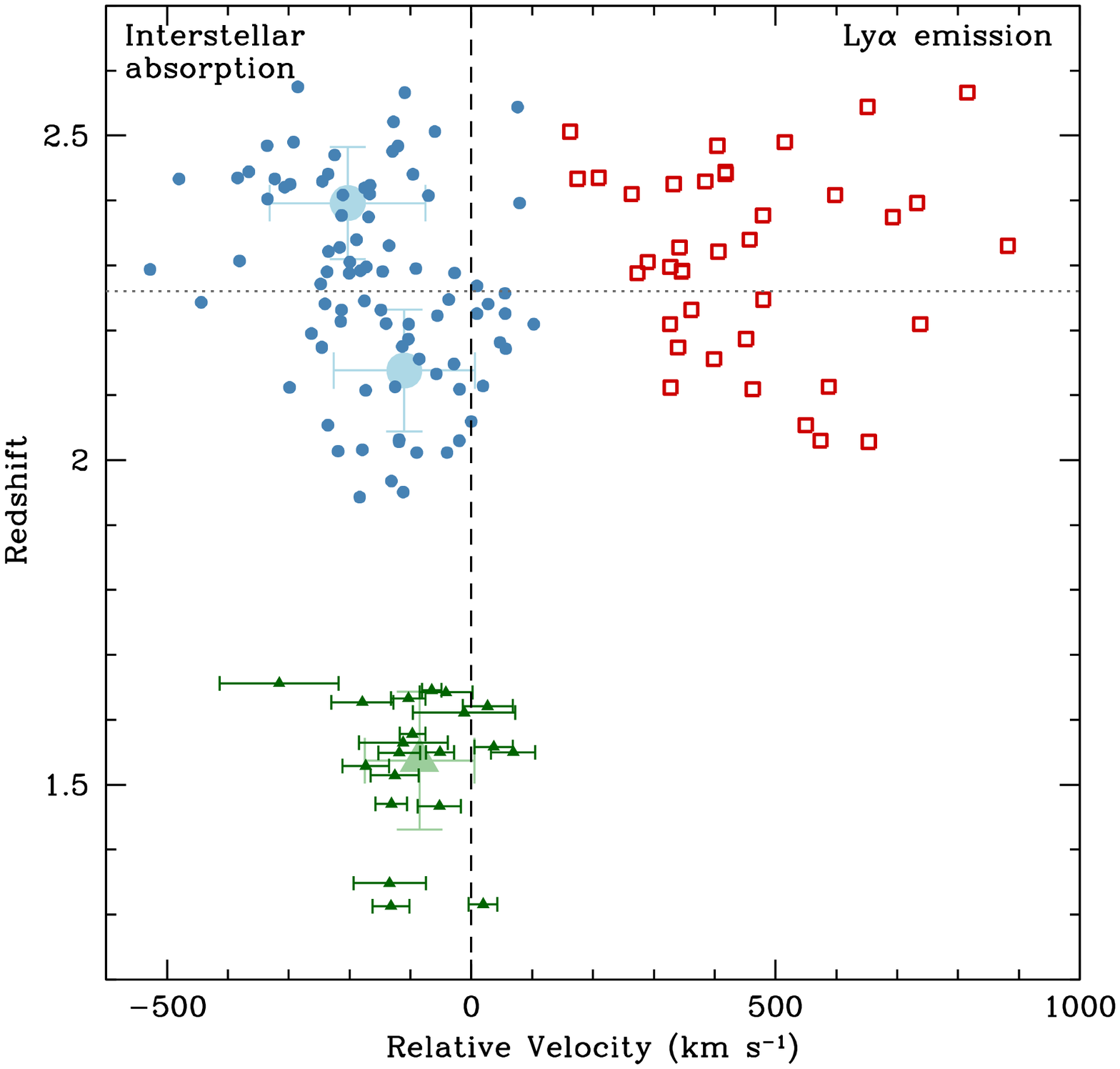}
\caption{Velocity centroids of interstellar absorption lines in galaxies at $z\sim2$ (blue circles, far-UV lines from \citealt{ses+10}) and $z\sim1.5$ (the current  \FeII\ $\lambda$2374 sample, green triangles with error bars) plotted against redshift. The velocity of \lya\ emission is also shown when present for the \ztwo\ sample (open red squares).  The horizonal dotted line divides the \ztwo\ sample into two redshift bins containing an equal number of galaxies, and the larger, lighter colored points show the mean velocity and redshift of the two bins of the \ztwo\ sample and the $z\sim1.5$ sample.  Velocities are relative to \Ha\ emission at \ztwo\ and \OII\ emission at $z\sim1.5$.}
\label{fig:deltav_z}
\end{figure}

The magnitude of the velocity offset $\Delta v_{\rm abs}$ is a factor of $\sim2$ smaller than the value of  $\Delta v = -166 \pm 130$ \kms\ (mean and standard deviation) determined by \citet{ses+10}.   We plot $\Delta v_{\rm abs}$ against redshift for the current sample and for the higher redshift sample of \citet{ses+10} in Figure \ref{fig:deltav_z} (see also Figure 2 of \citealt{ses+10}), clearly showing the smaller offset from systemic velocity in the lower redshift sample.   The horizontal dotted line divides the \citet{ses+10} sample into two redshift bins containing an equal number of galaxies, and the large light blue points show the mean velocity offset and redshift of each bin;  although the scatter is considerable, the average offset from systemic increases with increasing redshift, from $\Delta v = -110$ \kms\ at $z\sim2.1$ to $\Delta v = -203$ \kms\ at $z\sim2.4$.  This increase is apparently driven by a decreasing fraction of objects with velocities near zero, as well as a higher fraction with $\Delta v \lesssim -300$ \kms.  

Our lower redshift sample follows the same trend of decreasing velocity centroid with decreasing redshift.  A full explanation of this trend requires more careful accounting of selection effects, however; our galaxies are on average somewhat lower in mass than the \citet{ses+10} sample, since the magnitude limit of the survey reaches intrinsically fainter objects at lower redshifts.  Velocity offsets in the  \citet{ses+10} sample are measured from the far-UV interstellar absorption lines \CII\ $\lambda1334$, \ion{Si}{4} $\lambda1393$ and \SiII\ $\lambda1526$, whereas we use the near-UV \FeII\ lines. The kinematics of these different lines have not been directly compared, although \CII, \SiII\ and \FeII\ are expected to be similar as all of these transitions arise in cool outflowing gas, and \citet{ssp+03} find the kinematics of \ion{Si}{4} to be similar to those of the low ionization lines. Finally, a decrease in the blueshift of the velocity centroid does not necessarily correspond to a decrease in the outflow velocity, since the centroid may be influenced by the interstellar medium of the galaxy or by underlying emission near zero velocity.

For the 45 galaxies in the sample without \OII\ emission, we use Equation \ref{eq:zsys} to determine the systemic redshift.  Where possible, we use \FeII\ $\lambda$2374 to define the absorption redshift, and when that feature is unavailable, we use \FeII\ $\lambda$2587 (the choice of these two lines is motivated by the fact that they are least likely to be affected by emission filling, which may shift the wavelength to the blue).  In the 14 galaxies where these two features were not detected, we use the far-UV absorption redshift from prior Keck/LRIS spectroscopy \citep{ssp+04}.  The mean systemic redshift of the sample of 45 galaxies determined in this way is then $\langle z \rangle = 1.836 \pm 0.16$.  Given the possible trend of $\Delta v_{\rm abs}$ with redshift, there is some concern that the galaxies for which we apply Equation \ref{eq:zsys} lie at somewhat higher redshifts than the galaxies used to determine the equation;  however, the similar velocity offsets of our sample and the lower redshift half of the \citet{ses+10} sample suggest that this is not a large effect.  

It is desirable to verify systemic redshifts obtained in this way via measurement of weak stellar or nebular features in composite spectra; if systemic redshifts have been assigned correctly, these features should lie at zero velocity (e.g.\ \citealt{ssp+03}).  Unfortunately our spectra do not contain features ideal for this purpose. The strongest nebular emission line, \CIIsf\ $\lambda 2326$, is a blend of several closely separated transitions whose ratios depend on the electron density, making it unsuitable for precise redshift determinations.  The composite spectrum of all 96 galaxies does contain a weak detection of the \ion{C}{3} $\lambda2297$ stellar photospheric line (vacuum rest wavelength 2297.579 \AA), used by \citet{prs+02} to measure the systemic redshift of the lensed galaxy MS 1512-cB58.  The low S/N of this line makes it of limited use, but its velocity in the composite spectrum is $\Delta v = 19 \pm 78$ \kms, consistent with zero. We can also check the systemic redshifts by comparing the composite spectrum of the 51 galaxies with \OII-based redshifts with that of the 45 galaxies with redshifts determined using Equation \ref{eq:zsys}. For this test we compare the centroid and width of \FeII* $\lambda2626$, the strongest emission line appearing in both spectra; we find that the line centroids in the two spectra agree to within 20 \kms, while the line is $\sim0.8$ \AA\ (90 \kms) broader in the spectrum constructed using redshifts from Equation \ref{eq:zsys}. Such broadening is expected given the 87 \kms\ dispersion in $\Delta v_{\rm abs}$. These tests validate our use of Equation \ref{eq:zsys} to determine the systemic redshifts of galaxies without \OII\ emission. 

\section{Stellar Population Modeling}
\label{sec:sedfitting}
For 71 of the 96 galaxies in the sample, we determine the stellar mass, age, extinction and star formation rate through modeling of the broadband spectral energy distribution (SED).  We require photometry in the $K_s$ and/or {\it Spitzer} IRAC 3.6 or 4.5 \micron\ bands in order to perform the modeling, and most of the remaining 25 galaxies are not covered by this IR imaging; we do not expect these 25 galaxies to be significantly different from the rest of the sample. We model the SED as described by \citet{sse+05} and \citet{ess+06mass}; briefly, we use the \citet{bc03} stellar population models, and assume constant star formation, the \citet{cab+00} extinction law, and a \citet{c03} initial mass function.  Uncertainties in the fitted parameters are determined by Monte Carlo simulations which perturb the input photometry according to the photometric uncertainties and determine the best-fit SED for the perturbed colors; typical fractional uncertainties are $\sigma_x/\langle x \rangle=0.6$, 0.4, 0.5, and 0.2 in $E(B-V)$, age, SFR, and stellar mass, respectively.  Systematic uncertainties are discussed further in Section \ref{sec:compspecs}. We assume constant star formation histories because of the difficulty of obtaining robust constraints on the star formation history of galaxies from SED fitting alone; fortunately, the stellar mass is relatively insensitive to the assumed star formation history \citep{ess+06mass,rps+12}.

\begin{figure}[htbp]
\plotone{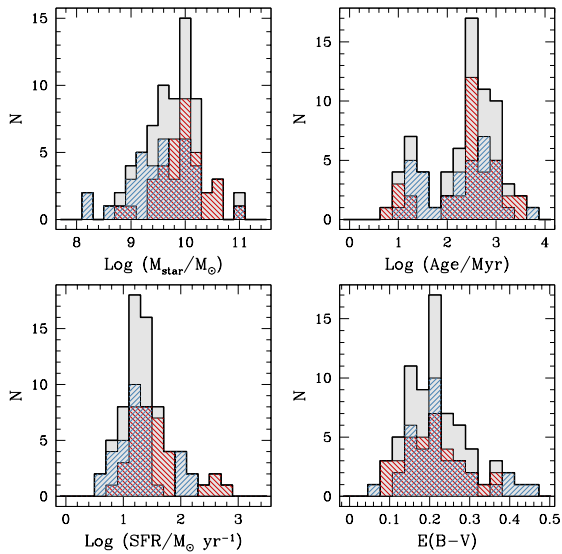}
\caption{Distributions of stellar population parameters for the 71 galaxies with SED fits are shown by the pale grey histograms.  We also divide the sample by redshift and show the distributions for the upper and lower halves:  the median redshift is $z_{\rm med}=1.65$, galaxies with $z<1.65$ are shown with blue histograms, and galaxies with $z>1.65$ in red.  Galaxies with smaller stellar masses and lower SFRs tend to be at lower redshifts because the magnitude-limited survey reaches intrinsically fainter objects at lower redshifts. Upper left panel: stellar mass; upper right: age; lower left: star formation rate; lower right: $E(B-V)$. }
\label{fig:sedhists}
\end{figure}

The range of stellar population parameters for the 71 galaxies is shown in Figure \ref{fig:sedhists}.  We find a mean (median) stellar mass $\langle M_{\star} \rangle = 1.1 \times 10^{10}$ \msun ($6.3\times10^9$ \msun), mean (median) reddening $\langle E(B-V) \rangle = 0.21$ (0.21), mean (median) age 550 Myr (360 Myr), and mean (median) SFR 54 \msunyr\ (22 \msunyr).  The degeneracies between age, extinction and star formation history involved in such modeling are well-known, as is the fact that the stellar mass is the most well-determined parameter (e.g.\ \citealt{pdf01,sse+05,rps+12}).  We discuss the relationships between the modeled quantities further in Section \ref{sec:composites}, where we construct composite spectra based on stellar population properties.

\section{Results from Individual Spectra}
\label{sec:individual}

\begin{figure*}[htbp]
\plotone{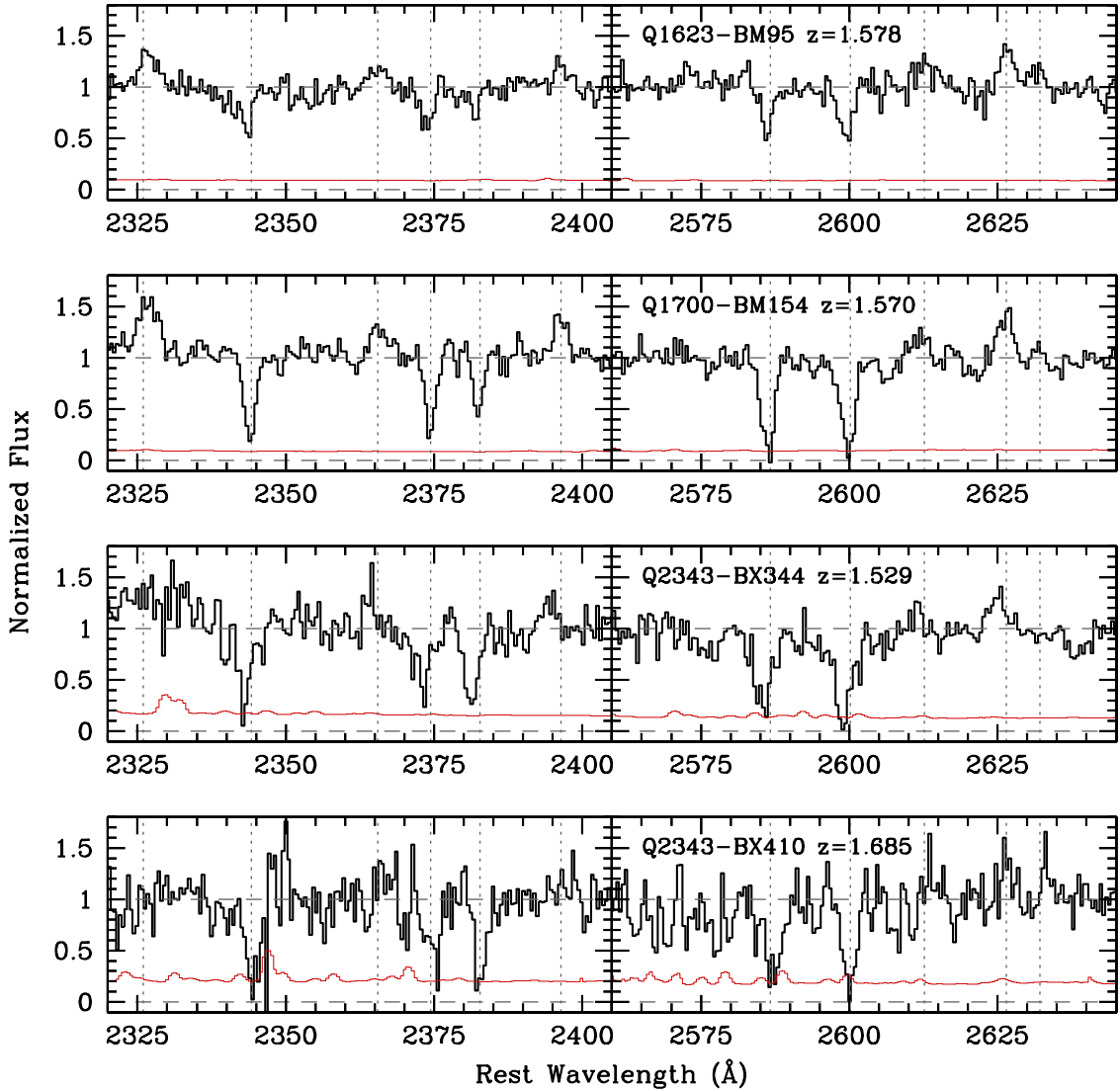}
\caption{Representative spectra of four individual galaxies, showing the complexes of \FeII\ absorption and emission.  All four galaxies have systemic redshifts from \OII\ emission. Features are marked by vertical dotted lines. Left panels, \CIIsf\ $\lambda 2326$ emission, \FeII\ $\lambda 2344$ absorption, \FeII* $\lambda 2365$ emission, \FeII\ $\lambda 2374$ absorption, \FeII\ $\lambda 2383$ absorption, and  \FeII* $\lambda 2396$ emission.  Right panels:   \FeII\ $\lambda 2587$ and  \FeII\ $\lambda 2600$ absorption,  \FeII* $\lambda 2612$,  $\lambda 2626$ and $\lambda 2632$ emission. }
\label{fig:ind_fespecs}
\end{figure*}

\begin{figure*}[htbp]
\plotone{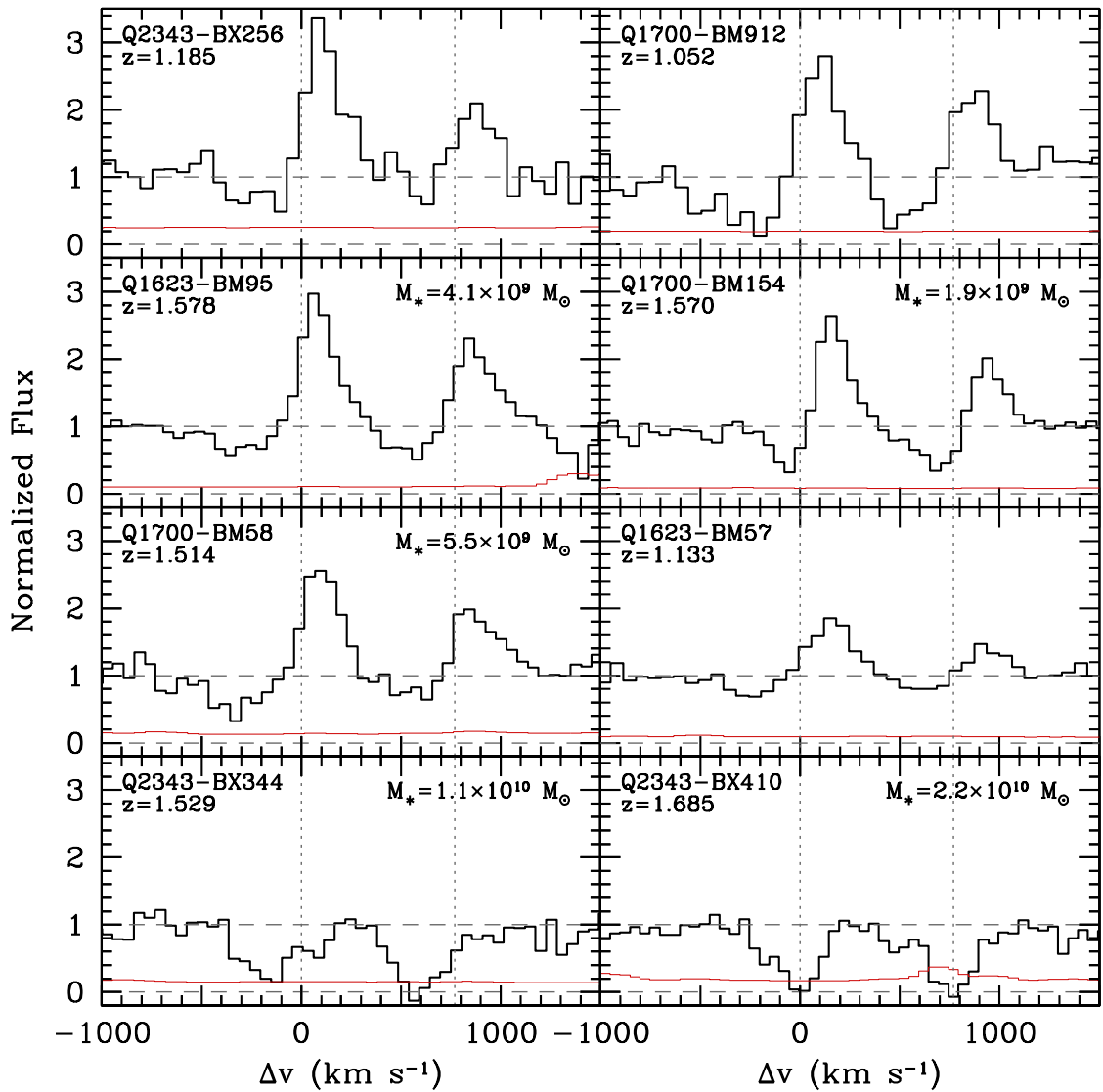}
\caption{The \MgII\ $\lambda\lambda$ 2796, 2803 profiles of eight of the galaxies in the sample, showing the range from strong emission to pure absorption.  Stellar masses are given when available, and galaxies with strong emission tend to be lower in mass.  All eight galaxies have systemic redshifts from \OII\ emission.}
\label{fig:mg2profiles}
\end{figure*}

Representative examples of the individual spectra are shown in Figures \ref{fig:ind_fespecs} and \ref{fig:mg2profiles}.  Figure \ref{fig:ind_fespecs}  shows the two complexes of \FeII\ absorption and emission lines, at $\sim2300$ and $\sim 2600$ \AA; the top two spectra are among the highest S/N in the sample, while the spectrum of Q2343-BX410 in the bottom panel is more representative of the typical data quality.  We also show the \MgII\ $\lambda\lambda$2796, 2803 profiles for these four galaxies and four additional objects in Figure \ref{fig:mg2profiles}.  While the \FeII\ absorption lines show some variation in their strengths, we see much greater diversity in the \MgII\ profiles, which range from strong emission with very little absorption (Q2343-BX256) to pure absorption (Q2343-BX410); the pattern is suggestive of that seen in the far-UV spectra of high redshift galaxies, in which \lya\ may appear in emission or absorption while the interstellar resonance lines appear in absorption. We study the relationships between the strengths of these lines and the properties of the galaxies in this section and in the section following.
	
\begin{figure}[htbp]
\plotone{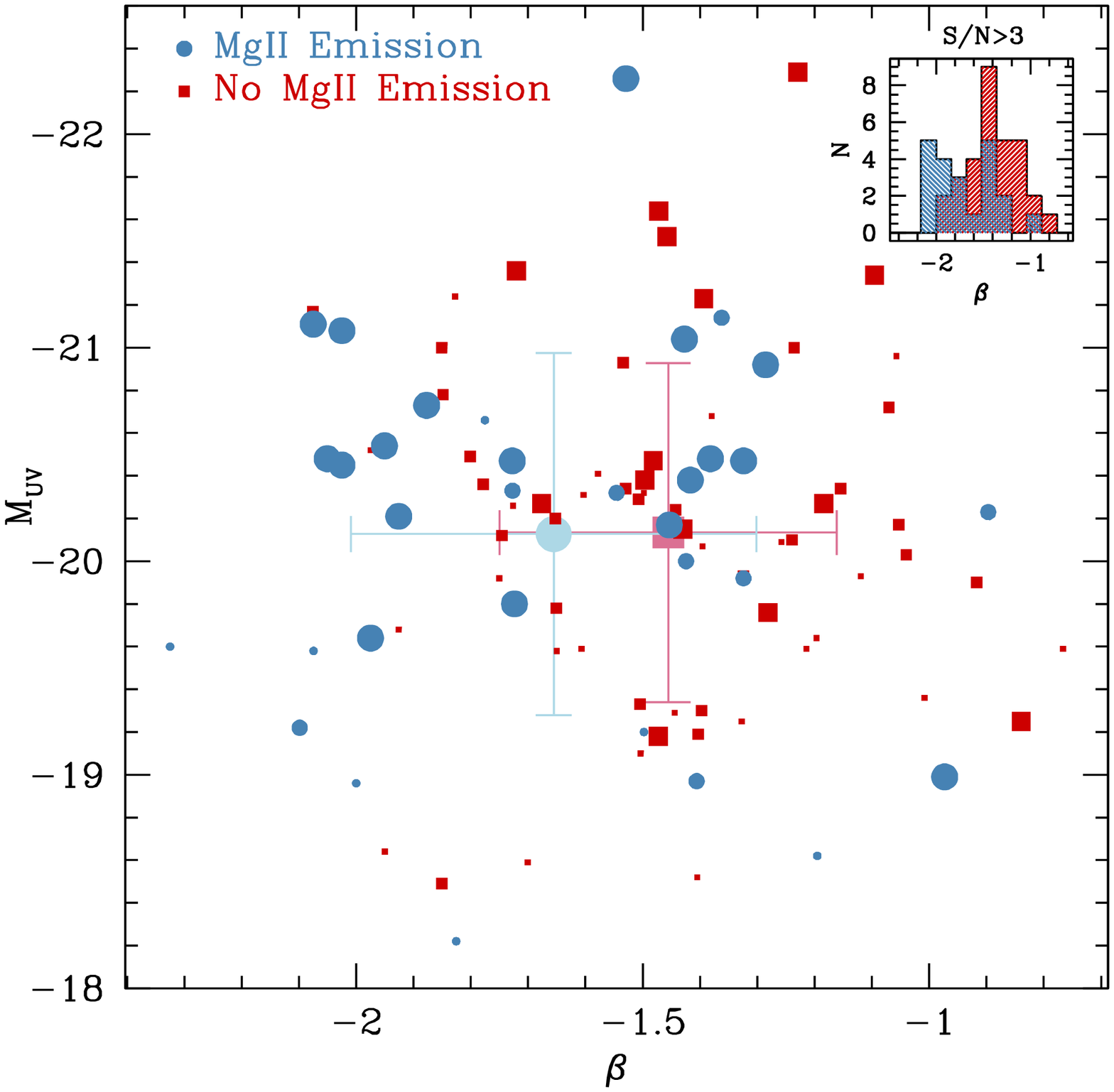}
\caption{Rest-frame UV slope $\beta$ vs.\ absolute UV magnitude $M_{\rm UV}$ for galaxies with \MgII\ emission (blue circles) and without (red squares).  The symbol size indicates the signal-to-noise ratio of the spectrum in the \MgII\ region:  galaxies plotted with large symbols have ${\rm S/N}>4.5$ per pixel, medium-sized symbols indicate galaxies with $2.5 < {\rm S/N} < 4.5$, and small symbols show galaxies with ${\rm S/N} < 2.5$. The large, lighter colored symbols show the average of the two samples, with error bars showing the standard deviations.  The inset at upper right shows the $\beta$ distributions of \MgII\ emitters (blue) and non-emitters (red) for spectra with ${\rm S/N}>3$.}
\label{fig:beta_mg2}
\end{figure}

The features covered by the spectra are listed in Table \ref{tab:lines}.  Many of these, particularly the weaker emission lines, are significantly detected only in the composite spectra discussed in the next section.  We measure the absorption and emission features by fitting a single Gaussian profile to each line using the IDL routine {\it mpfit.pro}; given a spectrum and the error spectrum containing the pixel-to-pixel uncertainties, this routine measures the center and equivalent width of each line and their associated uncertainties.  Although absorption lines associated with galactic outflows generally show asymmetric profiles (e.g.\ \citealt{psa+00, qpss09, qsp+10}), the S/N of the individual spectra in our sample is such that weak blue wings of the absorption lines are generally undetected and a Gaussian provides a good fit.  We do not attempt to fit outflowing and zero-velocity components separately, as some similar studies have done (e.g.\ \citealt{wcp+09, rwk+10}).
	
\begin{deluxetable*}{l l l l l l}
\tablewidth{0pt}
\tabletypesize{\footnotesize}
\tablecaption{Observed Absorption and Emission Features\label{tab:lines}}
\tablehead{
\colhead{Ion} &
\colhead{$\lambda_{\rm lab}$\tablenotemark{a}} & 
\colhead{$A$\tablenotemark{a}} & 
\colhead{$f$\tablenotemark{a}} & 
\colhead{Notes} \\
\colhead{} &
\colhead{(\AA)} &
\colhead{(s$^{-1}$)} &
\colhead{} &
\colhead{}
}
\startdata
\CIIsf          & 2326   & $4.43\times 10^{1}$ & ...&Emission; blend \\
\OII            & 2470.97        &  $2.12\times 10^{-2}$        & ...&Emission; blended with \OII\,$\lambda$2471\\
                & 2471.09 & $5.22\times 10^{-2}$        & ...&Emission; blended with \OII\,$\lambda$2470\\
\MgI            & 2852.96 & $4.91\times 10^{8}$         & 1.83&Absorption\\
\MgII           & 2796.35 & $2.60\times 10^{8}$         &0.6155 & Absorption and Emission\\
                & 2803.53 & $2.57\times 10^{8}$ & 0.3058& Absorption and Emission\\
Fe\,{\sc ii} & 2249.88 & ... & 0.00182&Absorption\\
                & 2260.78 & ...& 0.00244&Absorption\\
                & 2344.21       & $1.73\times 10^{8}$ & 0.114&Absorption\\
                & 2365.55 & $5.90\times 10^{7}$ & 0.0495 & Emission \\
                         & 2374.46      & $4.25\times 10^{7}$ & 0.0313&Absorption\\
                         & 2382.76      & $3.13\times 10^{8}$ & 0.320&Absorption\\
                         & 2396.36 & $2.67\times 10^{8}$ & 0.279 &Emission\\
                         & 2586.65      & $8.94\times 10^{7}$  & 0.0691&Absorption\\ 
                         & 2600.17      & $2.35\times 10^{8}$ & 0.239&Absorption \\ 
                         & 2612.11 & $1.23\times 10^{8}$ & 0.122 &Emission\\
                         & 2626.45 & $3.41\times 10^{7}$ & 0.0455 &Emission\\
                         & 2632.11 & $6.21\times 10^{7}$ & 0.087&Emission
\enddata
\tablenotetext{a}{Vacuum wavelengths, Einstein $A$-values, and oscillator strengths from \citet{morton03} when available and from the NIST Atomic Spectra Database otherwise.}
\end{deluxetable*}

\subsection{Rest-frame UV Color and \MgII\ Emission}
\label{sec:mg2uvcolor}
We have seen from the spectra shown in Figure \ref{fig:mg2profiles} that the galaxies in our sample show a wide range in  \MgII\ $\lambda\lambda$2796, 2803 profiles, from strong emission to pure absorption.  In order to separate the sample into galaxies with \MgII\ emission and galaxies without, we classify as \MgII\ emitters those galaxies with at least two adjacent pixels at least 1.5$\sigma$ above the continuum level in either of the two \MgII\ transitions.  With this classification, 33 of the 96 galaxies in the sample show \MgII\ emission, while 63 do not. The classification clearly depends on the S/N of the individual spectra as well as the intrinsic strength of \MgII\ emission.

In Figure  \ref{fig:beta_mg2} we plot the rest-frame UV continuum slope $\beta$ ($f_{\lambda} \propto \lambda^\beta$) against $M_{\rm UV}$, the absolute magnitude at 1900 \AA, for galaxies with and without \MgII\ emission, using larger symbols for galaxies with higher S/N in the \MgII\ region of the spectrum.  We measure $\beta$ from the $G-{\cal R}$ color, which is uncorrelated with redshift for the galaxies in our sample.  At the sample's mean redshift of $\langle z \rangle = 1.6$, the $G$ filter is centered at a rest wavelength of 1860 \AA, and the ${\cal R}$ filter lies at 2665 \AA.  We calculate the absolute magnitude $M_{\rm UV}$ at $\sim1900$ \AA, using the $G$ magnitude for $1.4<z<1.8$, and composite magnitudes $m_{\rm UG}$ and $m_{\rm G{\cal R}}$ for $z<1.4$ and $z>1.8$ respectively.  Composite magnitudes are determined by averaging the flux in the two relevant filters.

It is clear that galaxies with significant \MgII\ emission tend to have bluer UV slopes. Quantitatively, 17 of the 33 galaxies with \MgII\ emission are found in the bluest third of the sample, with $\beta<-1.7$, while the remaining 16 are equally distributed between the two redder thirds. A two-sample K-S test finds that the probability that the \MgII\ emitters and non-emitters are drawn from the same $\beta$ distribution is 0.025. This conclusion is somewhat strengthened when we consider only the 52 galaxies with ${\rm S/N}>3$ in the \MgII\ region; among this subsample, the 21 \MgII\ emitters have $\langle \beta \rangle = -1.69 \pm 0.35$ (where 0.35 is the standard deviation), while the 31 non-emitters have $\langle \beta \rangle = -1.40 \pm 0.27$.  Histograms of these two subsamples are shown in the inset at the upper right of Figure \ref{fig:beta_mg2}. Repeating the K-S test indicates that the probability that the two samples are drawn from the same distribution is 0.0097.  In contrast to the $\beta$ distributions, the emitting and non-emitting samples are similarly distributed in UV luminosity.

\MgII\ emission was also identified in a sample of $\sim1400$ $z\sim1.4$ galaxies by \citet{wcp+09}, using criteria which select an excess of emission above the continuum level in a window between the two \MgII\ absorption lines.  The excess emission objects selected in this way constituted $\sim 4$\% of the sample, and were attributed to narrow-line AGN.   \citet{wcp+09} found that these excess emission objects had bluer $U-B$ colors and (in contrast to our results) were more luminous.  They also found that the excess emission was usually at the systemic redshift and not always accompanied by absorption. 

Our fraction of \MgII\ emitters is considerably higher than that of \citet{wcp+09}, although the different selection criteria make a direct comparison difficult.  We attribute the emission to resonant scattering in outflowing gas; as shown in Figure \ref{fig:mg2profiles}, the emission is generally redshifted and accompanied by blueshifted absorption, the classic P Cygni profile of an outflow.  The P Cyngi profile is also seen in \lya\ emission and absorption in galaxies at high redshift, and the relationship between UV color and \lya\ emission is  similar to the relationship between UV color and \MgII\ emission: galaxies with stronger \lya\ emission are observed to have bluer UV slopes \citep{ssp+03, jse12}.  We discuss the relationship between \MgII\ emission and galaxy properties further in Section \ref{sec:composites}.

\subsection{Relative Velocities and Equivalent Widths of Absorption and Emission Lines}
\label{sec:velw_indspec}

As discussed in Section \ref{sec:intro}, most of the \FeII\ absorption lines are coupled to fine structure transitions, such that a photon absorbed at (for example) 2587 \AA\ can be emitted in the \FeII* $\lambda$2612 or $\lambda$2632 emission lines.  The exception is \FeII\ $\lambda 2383$; the only allowed transition for a photon absorbed in this line is a return to the ground state.  Thus, as discussed in detail by \citet{pkr11}, one may expect significant rescattered emission to ``fill in" the absorption in this and other resonant transitions; one may also expect the filling to be greater when the photon has no other means to return to a lower energy level, i.e.\ for transitions such as \FeII\ $\lambda 2383$ which are not coupled to fine structure emission lines.  Evidence for this emission filling may be seen in the form of blueshifted velocity centroids, since the scattered emission is expected to occur near systemic velocity, or in a decrease in absorption equivalent width.  Either of these effects may influence conclusions drawn regarding the kinematics, optical depth and covering fraction of outflowing gas.  The remainder of this section is dedicated to assessing the effect of this emission filling on measurements of velocity centroids and equivalent widths.
	
\begin{figure*}[htbp]
\plotone{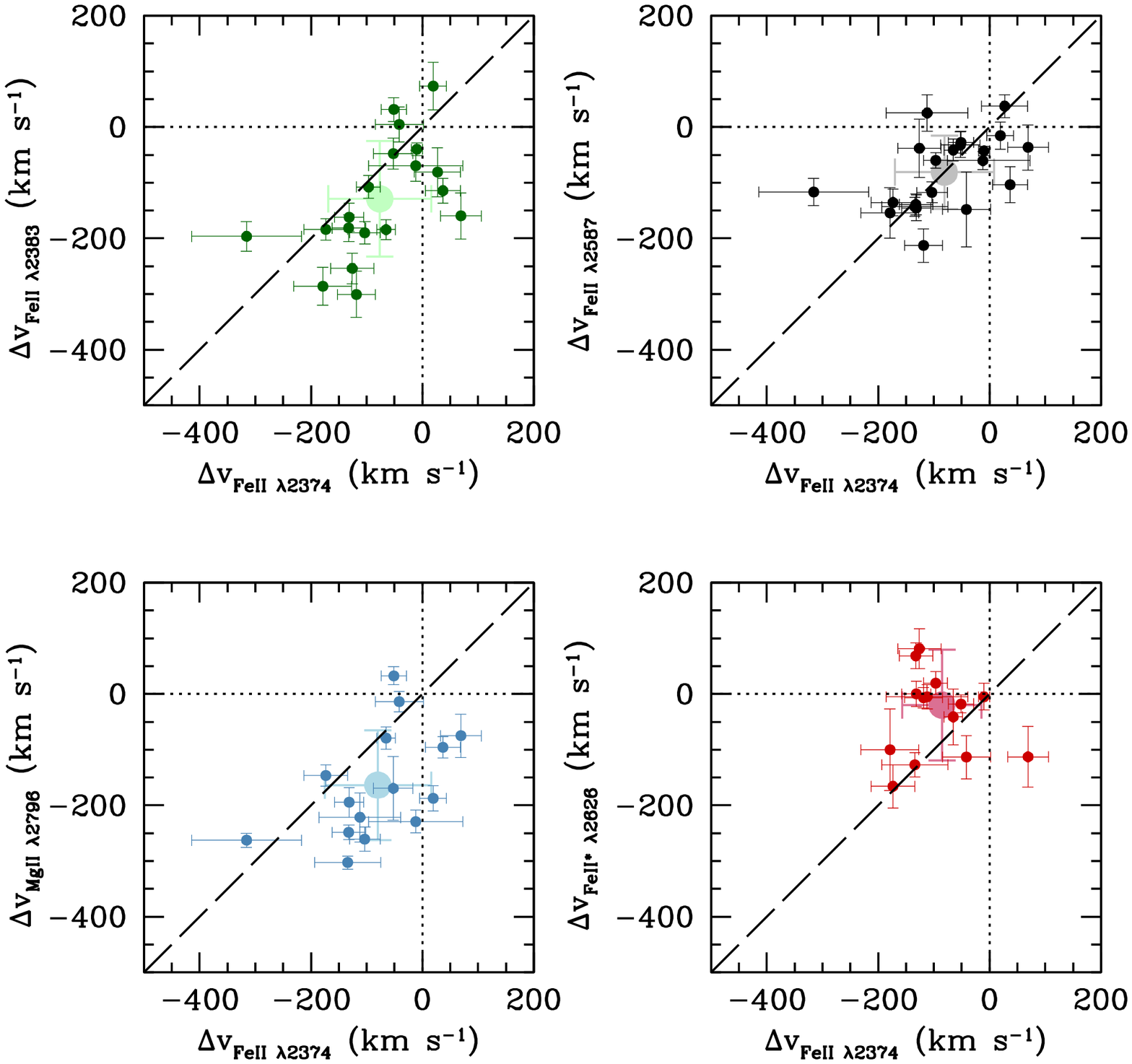}
\caption{The velocities of \FeII\ $\lambda$2383 absorption (upper left),  \FeII\ $\lambda$2587 absorption (upper right),  \MgII\ $\lambda$2796 absorption (lower left) and  \FeII* $\lambda$2626 emission (lower right), each plotted against the velocity of \FeII\ $\lambda$2374 absorption.  Only objects with systemic redshifts from \OII\ emission are included, and the large, light-colored points show the average value for each line.}
\label{fig:ind_velcomp}
\end{figure*}		
	
In Figure \ref{fig:ind_velcomp} we compare the velocities with respect to systemic for various transitions, as measured by the center of the Gaussian fit. We plot only galaxies with velocity uncertainties $\sigma_{\Delta v}<100$ \kms\ (an absolute uncertainty is preferable to some threshold in $\sigma_{\Delta v}/\Delta v$ in order to avoid biasing the sample away from galaxies with $\Delta v$ near zero).  Because this measurement relies on a precise determination of the systemic redshift for each object, we include only galaxies for which we have determined $z_{\rm sys}$ from \OII\ emission.  We plot the velocity centroids of \FeII\ $\lambda 2383$ and \FeII\ $\lambda 2587$ absorption, \MgII\ $\lambda$2796 absorption, and \FeII* $\lambda 2626$ emission, all against that of \FeII\ $\lambda 2374$.  As discussed above, we may expect \FeII\ $\lambda 2383$ to show a larger blueshift than \FeII\ $\lambda 2374$ or \FeII\ $\lambda 2587$, and this is indeed observed in the upper two panels of Figure \ref{fig:ind_velcomp}: on average, \FeII\ $\lambda 2374$ and \FeII\ $\lambda 2587$ show identical blueshifts of $-81$ \kms, while the average relative shifts of \FeII\ $\lambda 2374$ and \FeII\ $\lambda 2383$ are $-77$ and $-129$ \kms\ respectively (the two averages for \FeII\ $\lambda 2374$ are not identical because we include only galaxies with $\sigma_{\Delta v}<100$ \kms\ for both lines in each plot, and the two subsamples are slightly different).
	
We also compare the relative velocity shifts of \FeII\ $\lambda 2374$ and \MgII\ $\lambda$2796, in the lower left panel of Figure \ref{fig:ind_velcomp}.  Because both \FeII\ $\lambda 2383$ and \MgII\ $\lambda$2796 are resonance transitions coupled only to the ground state, we might expect emission filling to affect them similarly; however, as also suggested by the diversity of \MgII\ profiles relative to \FeII\ in Figures \ref{fig:ind_fespecs} and \ref{fig:mg2profiles}, the \MgII\ transition shows stronger evidence for the presence of emission.  The average blueshift of \MgII\ is $-164$ \kms, compared to the $-129$ \kms\  of \FeII\ $\lambda 2383$.  In the case of both lines, the velocity comparison shows that only a fraction of the sample shows evidence of emission underlying the absorption; about $\sim1/3$ of the \FeII\ $\lambda 2383$ sample is significantly offset from the \FeII\ $\lambda 2374$ absorption, while the fraction of the sample showing an offset in \MgII\ $\lambda$2796 is higher at $\sim 2/3$.

Finally, we compare the velocity centroids of  \FeII\ $\lambda 2374$ with those of the strongest \FeII\ fine structure emission line, \FeII* $\lambda 2626$. We discuss the origin of the fine structure emission lines in more detail in the following sections; for now, we note that \citet{rpm+11} have proposed that \FeII* emission arises from photon scattering in the outflowing gas and should be observed at or near systemic velocity, assuming that the outflow is approximately isotropic and has little extinction. With a mean velocity centroid of $\Delta v_{2626} = -20$ \kms, our observations are in general agreement with this scenario.

We turn next to a comparison of the equivalent widths of the \FeII\ absorption lines, measured by integrating the best-fit Gaussian profiles. For an optically thin transition, the equivalent width $W$ of an absorption line depends on the covering fraction and column density $N$ of the ion and the oscillator strength of the transition $f$ such that $W \propto N\lambda^2 f$.  When the transition becomes optically thick, the equivalent width depends instead on the velocity dispersion and covering fraction of the absorbing gas, and we then expect (in the absence of emission filling) different transitions of the same ion to have the same equivalent width.  It is therefore instructive to compare the equivalent widths of some of the observed \FeII\ absorption lines. 
	
\begin{figure}[htbp]
\plotone{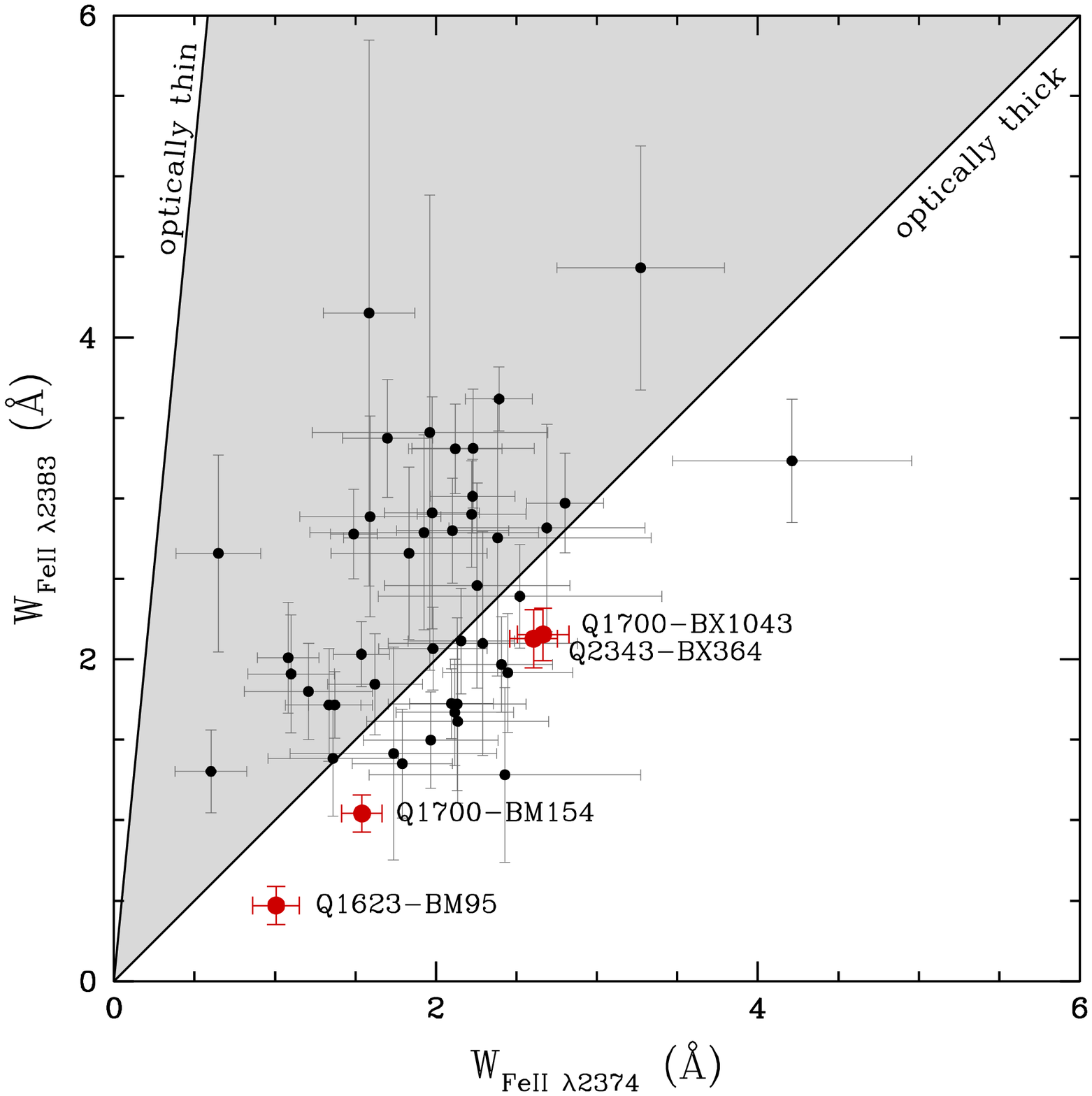}
\caption{A comparison of the equivalent widths of \FeII\ $\lambda$2374 and \FeII\ $\lambda$2383.  The shaded area shows the allowed region, between optically thin and thick transitions; the four marked objects lie significantly outside this region, and are discussed in the text.}
\label{fig:ind_ewcomp}
\end{figure}	

In Figure \ref{fig:ind_ewcomp} we compare the equivalent widths of \FeII\ $\lambda$2374 and \FeII\ $\lambda$2383.  With oscillator strengths of $f=0.0313$ and $f=0.320$ respectively, these two lines should have very different equivalent widths in the optically thin case, with $W_{\rm Fe II \lambda 2383}/W_{\rm Fe II  \lambda 2374} = 10.3$.  As the absorbing gas becomes optically thick, we expect $W_{\rm Fe II \lambda 2383}/W_{\rm Fe II  \lambda 2374} \approx 1$.  An additional effect is the possible presence of emission filling: as discussed above, the two transitions  differ in that \FeII\ $\lambda 2374$ is coupled to a fine structure transition, while \FeII\ $\lambda 2383$ is not, and we therefore expect that \FeII\ $\lambda 2383$ is more likely to be affected by the presence of emission underlying the absorption line.  Such underlying emission would lead to a decrease in equivalent width, producing, in the optically thick case, otherwise unphysical line ratios of $W_{\rm Fe II \lambda 2383}/W_{\rm Fe II  \lambda 2374} < 1$.

Figure \ref{fig:ind_ewcomp} shows that in most cases $W_{\rm Fe II \lambda 2383}/W_{\rm Fe II  \lambda 2374}$ is consistent with unity or falls between the saturated and unsaturated limits.  However, there are four objects (Q1623-BM95, Q1700-BM154, Q1700-BX1043 and Q2343-BX364, shown with large red circles in Figure \ref{fig:ind_ewcomp}) for which the equivalent width ratio of the two lines can only be explained by a decreased  \FeII\ $\lambda 2383$  equivalent width due to underlying emission.  Note that this represents only a lower limit on the number of objects affected by emission filling; if the intrinsic line ratio falls between the unsaturated and saturated limits, as is the case for many of the galaxies in the sample, the equivalent width comparison cannot distinguish between emission filling and an increase in optical depth.  There are also several additional objects that fall in the unshaded region with lower significance, so there may be additional galaxies in the sample with $W_{\rm Fe II \lambda 2383}/W_{\rm Fe II  \lambda 2374} < 1$.

As shown in Figure \ref{fig:mg2profiles}, Q1623-BM95 and Q1700-BM154 also have strong \MgII\ emission; Q1700-BX1043 and Q2343-BX364 (not shown in Figure \ref{fig:mg2profiles}) have lower S/N in the \MgII\ region, making the presence of emission more difficult to assess, but using the criterion described in Section \ref{sec:mg2uvcolor} BX1043 is also an \MgII\ emitter while BX364 is not.  We also note that the stellar masses of Q1623-BM95, Q1700-BM154, Q1700-BX1043 and Q2343-BX364 are $4.1\times10^9$ \msun, $1.9\times10^9$ \msun, $3.9\times10^9$ \msun\ and $4.5\times10^9$ \msun\ respectively, all lower than the average stellar mass of $1.1\times10^{10}$ \msun.  This small sample suggests that galaxies with strong \MgII\ emission are more likely to exhibit emission filling in the \FeII\ transitions as well, and that such emission filling may be more likely in low mass galaxies.  The fraction of galaxies significantly affected by this emission is relatively small, however, particularly for the \FeII\ lines.  
\section{Results from Composite Spectra}
\label{sec:composites}
In order to examine weak emission and absorption lines and study the variation of spectral features with galaxy properties more closely, we construct composite spectra based on the modeled stellar population properties and on the presence or absence of \MgII\ emission.  We also create a composite spectrum of all 96 galaxies in the sample.

The stacked spectra represent pixel-by-pixel averages of the normalized individual spectra, created with 3$\sigma$ rejection. The mean and  standard deviation of the flux values at a given wavelength step were calculated, pixels with flux values $\geq 3\sigma$ were rejected, and the mean flux value of the remaining pixels was output as the final composite spectrum.  Because this was done for each pixel, the individual spectra contributing to the mean flux at a given pixel vary. The error spectrum for each composite was computed by adding in quadrature the error spectra for the individual galaxies contributing to the mean flux at each wavelength step. The final stacked spectra have 0.25\,\AA\ per pixel in the rest frame.

\begin{figure*}[htbp]
\plotone{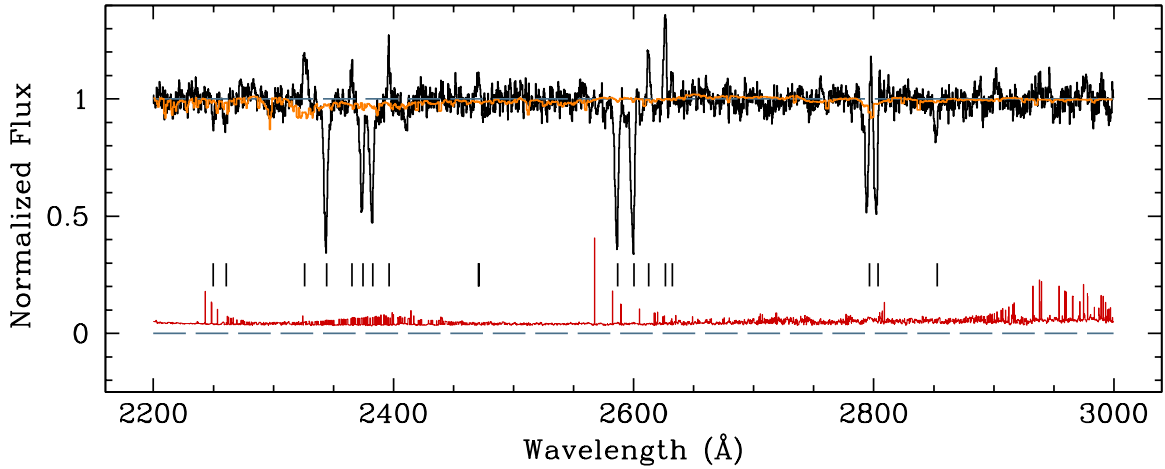}
\caption{The composite spectrum of all 96 galaxies in the sample.  The thin red line shows the 1$\sigma$ error spectrum, and features listed in Table \ref{tab:lines} are marked with vertical black lines.  The overplotted orange line is the stellar spectrum of a solar metallicity, 100 Myr-old galaxy with constant star formation as predicted by Starburst99, as discussed in the text. The mean S/N per pixel of the composite spectrum is 21.}
\label{fig:all_fullspec}
\end{figure*}

The composite spectrum of the full sample of 96 galaxies is shown in Figure \ref{fig:all_fullspec}, with the features listed in Table \ref{tab:lines} marked with vertical black lines.  In addition to the strong \FeII\ and \MgII\ absorption lines, we clearly detect five \FeII* emission lines and the broad blend of nebular \CIIsf\ emission at 2326 \AA.  Figure \ref{fig:all_fullspec} also shows (overplotted in orange) the synthetic, normalized stellar spectrum of an integrated stellar population with constant star formation, smoothed to a resolution of $\sim180$ \kms\ (FWHM) to match the resolution of our observations.  The spectrum was produced by the stellar synthesis code Starburst99 \citep{sb99}, and incorporates the library of theoretical UV spectra of massive stars described by \citet{lob+10}.  The stellar spectrum assumes solar metallicity, an age of 100 Myr (the UV spectrum stabilizes after $\sim50$ Myr under conditions of constant star formation) and a \citet{k01} initial mass function.  This spectrum shows that the stellar contribution to the features of interest for the current study is negligible; for constant star formation, the stellar continuum at near-UV wavelengths is dominated by stars too hot to have strong \FeII\ or \MgII\ absorption.  

\subsection{Composite Spectra Based on Stellar Population Properties}
\label{sec:compspecs}
As described in Section \ref{sec:sedfitting}, we model the broadband SEDs of 71 of the 96 galaxies in the sample to determine their stellar masses, ages, star formation rates and reddening.  We divide the 71 galaxies into three bins based on each of these four parameters and construct composite spectra of the galaxies in the highest and lowest bins, resulting in 8 composite spectra, each constructed from 24 individual spectra.  The mean stellar population properties of the galaxies in each composite are listed in Table \ref{tab:sedcomps}, along with the mean S/N per pixel of each composite spectrum.
	      	            
\begin{deluxetable*}{l c c c c c c}
\tablewidth{0pt}
\tabletypesize{\footnotesize}
\tablecaption{Mean Stellar Population Properties of Composite Spectra\label{tab:sedcomps}}
\tablehead{
\colhead{Composite} &
\colhead{N} & 
\colhead{S/N} &
\colhead{Stellar Mass} & 
\colhead{E(B-V)} &
\colhead{SFR} &
\colhead{Age}\\
\colhead{} &
\colhead{} &
\colhead{(per pixel)} &
\colhead{($10^9$ \msun)} &
\colhead{} &
\colhead{(\msunyr)} &
\colhead{(Myr)}
}
\startdata
High Mass & 24 & 14 & 24.3 & 0.20 & \phantom{9}28 & 1119\\
Low Mass & 24 & 11 & \phantom{9}1.9 & 0.25 & \phantom{9}54 & \phantom{9}133\\
High E(B-V) & 24 & 12 & 12.0 & 0.31 & 118 & \phantom{9}363\\
Low E(B-V) & 24 & 15 &\phantom{9}7.7 & 0.14 & \phantom{9}18 & \phantom{9}589\\
High SFR & 24 & 15 & \phantom{9}9.5 & 0.28 & 128 & \phantom{9}205\\
Low SFR & 24 & 11 & \phantom{9}8.1 & 0.17 & \phantom{9}10 & \phantom{9}768\\
High Age & 24 & 11 & 21.7 & 0.18 & \phantom{9}16 & 1260\\
Low Age & 24 & 12 & \phantom{9}2.7 & 0.28 & 118 & \phantom{99}55

\enddata
\end{deluxetable*}
	       
There is, of course, considerable overlap between the composites because of covariance in the fitted parameters.  We quantify this overlap in Table \ref{tab:overlap}, in order to clarify the relationships between the stellar population properties considered.  This table shows the number of objects in common between any two composite spectra.   The largest overlap is between mass and age; because the models assume constant star formation, the stellar mass is simply the product of the age and star formation rate.  As shown in the table, the mass and age composites have 18/24 objects in common in the massive, old sample and 16/24 in the low mass and young sample.  

\begin{deluxetable*}{l c c c c c c c c}
\tablewidth{0pt}
\tabletypesize{\footnotesize}
\tablecaption{Overlap Between Composite Spectra\tablenotemark{a}\label{tab:overlap}}
\tablehead{
\colhead{} &
\colhead{Low} & 
\colhead{High} & 
\colhead{Low} &
\colhead{High} &
\colhead{Low} &
\colhead{High} &
\colhead{Low} &
\colhead{High} \\
\colhead{} &
\colhead{Mass} & 
\colhead{Mass} & 
\colhead{Age} &
\colhead{Age} &
\colhead{E(B-V)} &
\colhead{E(B-V)} &
\colhead{SFR} &
\colhead{SFR}
}
\startdata
Low Mass  &     ...  &  ...  &  18  &   \phantom{0}1  & \phantom{0}8  & 10  &   \phantom{0}9  & \phantom{0}9\\
High Mass  &    ...  &  ...  &  \phantom{0}0  & 16  &   \phantom{0}9  & \phantom{0}6  & \phantom{0}6  & \phantom{0}7\\
Low Age  &      18  & \phantom{0}0  &   ...  &  ...  &  \phantom{0}5  & 15  & \phantom{0}3  &   15\\
High Age  &     \phantom{0}1  & 16  &   ...  &  ...  &  12  &   \phantom{0}5  & 13  & \phantom{0}2\\
Low E(B-V)  &   \phantom{0}8  & \phantom{0}9  & \phantom{0}5  & 12  &   ...  &  ...  &  13  & \phantom{0}3\\
High E(B-V)  &  10  &   \phantom{0}6  & 15  & \phantom{0}5  &   ...  &  ...  &  \phantom{0}2  &         15\\
Low SFR  &      \phantom{0}9  & \phantom{0}6  & \phantom{0}3  & 13  &   13  & \phantom{0}2  &   ...  &  ...\\
High SFR  &     \phantom{0}9  & \phantom{0}7  & 15  & \phantom{0}2  & \phantom{0}3  &   15  &   ...  &...
\enddata
\tablenotetext{a}{Number of galaxies in common between any two composite
  spectra. Each composite represents the average of 24 galaxies.}
\end{deluxetable*}

Recent results indicate that the use of the \citet{cab+00} extinction law for galaxies with young best-fit ages overestimates their star formation rates with respect to other indicators; these galaxies may have less extinction than is suggested by their red UV slopes and be better modeled with a steeper, SMC-like extinction law \citep{rep+10,rps+12}.  The result is that $E(B-V)$ and the SFR may be overestimated for the youngest galaxies in the sample (and because of the covariance between extinction and age in the modeling, the ages of these objects are also uncertain).  This systematic uncertainty primarily affects galaxies with best-fit ages (calculated with models assuming the Calzetti extinction law) less than 50 Myr, of which there are 16 in our sample of 71 objects with SED fits. \citet{rps+12} find that restricting the ages to be at least 50 Myr (approximately equal to the dynamical time) and using the SMC extinction law for such objects brings the SED-determined SFRs into agreement with SFRs determined from the combined unobscured UV + IR luminosities. The effect of these changes can be significant; the typical result is to decrease the best-fit $E(B-V)$ by $\sim0.2$ mag, and reduce the SFR by a factor of $\sim8$.  Because of these uncertainties, in the following discussions we give greatest weight to the composite spectra based on stellar mass.

\begin{figure*}[htbp]
\epsscale{0.7}
\plotone{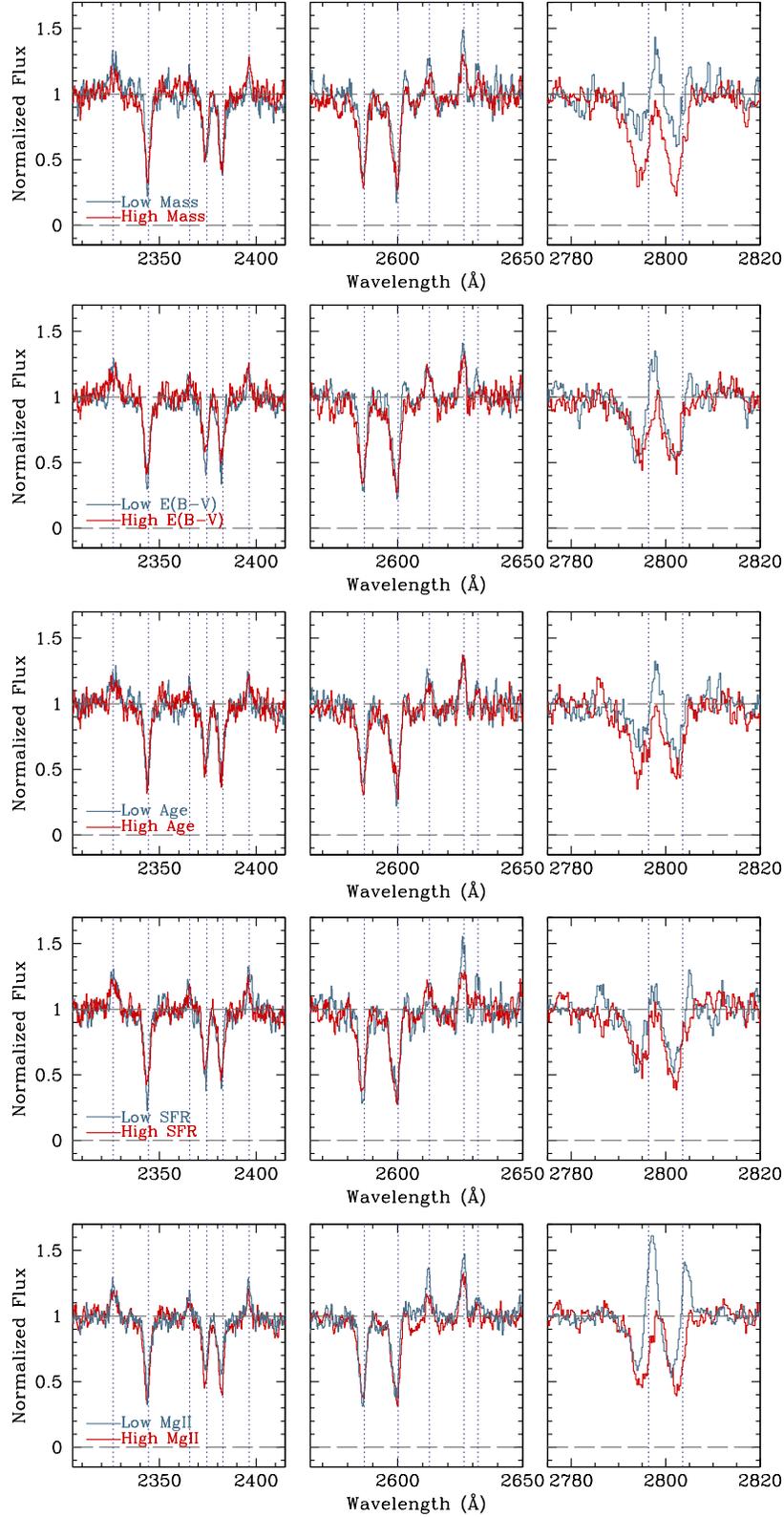}
\caption{Composite spectra constructed as a function of stellar population properties and the presence or absence of \MgII\ emission.  The spectra focus on the \FeII\ transitions at $\sim2350$ \AA\ in the left panels, the \FeII\ transitions at $\sim2600$ \AA\ in the middle panels, and \MgII\ $\lambda\lambda$2796, 2803 in the right panels.  From top to bottom, we show spectra binned by stellar mass, $E(B-V)$, age, star formation rate, and \MgII\ emission.  Red lines show the high mass, dust, age and SFR subsets, and the galaxies without \MgII\ emission.  Blue lines show the low mass, dust, age and SFR subsets, and the galaxies with \MgII\ emission.}
\label{fig:allcomposites}
\end{figure*}

We have also constructed composite spectra of the 33 galaxies with and 63 galaxies without \MgII\ emission, determined as described in Section \ref{sec:mg2uvcolor}.  Table \ref{tab:sedcomps} does not report mean stellar population properties for these composites, since not all of the galaxies have photometry sufficient for SED modeling; however, the composites with and without \MgII\ emission have mean S/N per pixel of 16.  The composite spectra based on the stellar population properties and on \MgII\ emission are shown in Figure \ref{fig:allcomposites}.

Equivalent widths and velocity centroids are measured by direct integration of the flux above or below the continuum level, with $1\sigma$ uncertainties determined by Monte Carlo simulations in which we perturb each pixel of the original composite spectrum by an amount dependent on its uncertainty 5000 times to create 5000 artificial spectra. The resulting velocity measurements are flux-weighted centroids, measured over a typical range of $\sim900$ \kms\  for the absorption lines and $\sim500$ \kms\ for the emission lines. The uncertainties adopted for \Wo\ and velocity are the standard deviation of the 5000 measurements of these values made for each feature. Measurements of the equivalent widths and velocity centroids of the absorption features are given in Table \ref{tab:abs}, and of the emission features in Table \ref{tab:em}.

Although we measured an average velocity shift of $-85$ \kms\ relative to systemic for the \FeII\ $\lambda 2374$ absorption line and then applied this shift to galaxies without \OII\ detections in order to construct the composites, the velocity centroid of FeII\ $\lambda 2374$ in the composite spectrum of all galaxies is  $-120$ \kms. This difference is due to the asymmetric absorption line profiles of the higher S/N composite, in which an extended blue tail of absorption shifts the centroids to larger velocities when they are measured with direct integration rather than Gaussian fits which assume symmetry.  The peak optical depth of the line is consistent with the $-85$ \kms\ measurements from individual spectra.
	     
\begin{figure*}[b]
\plotone{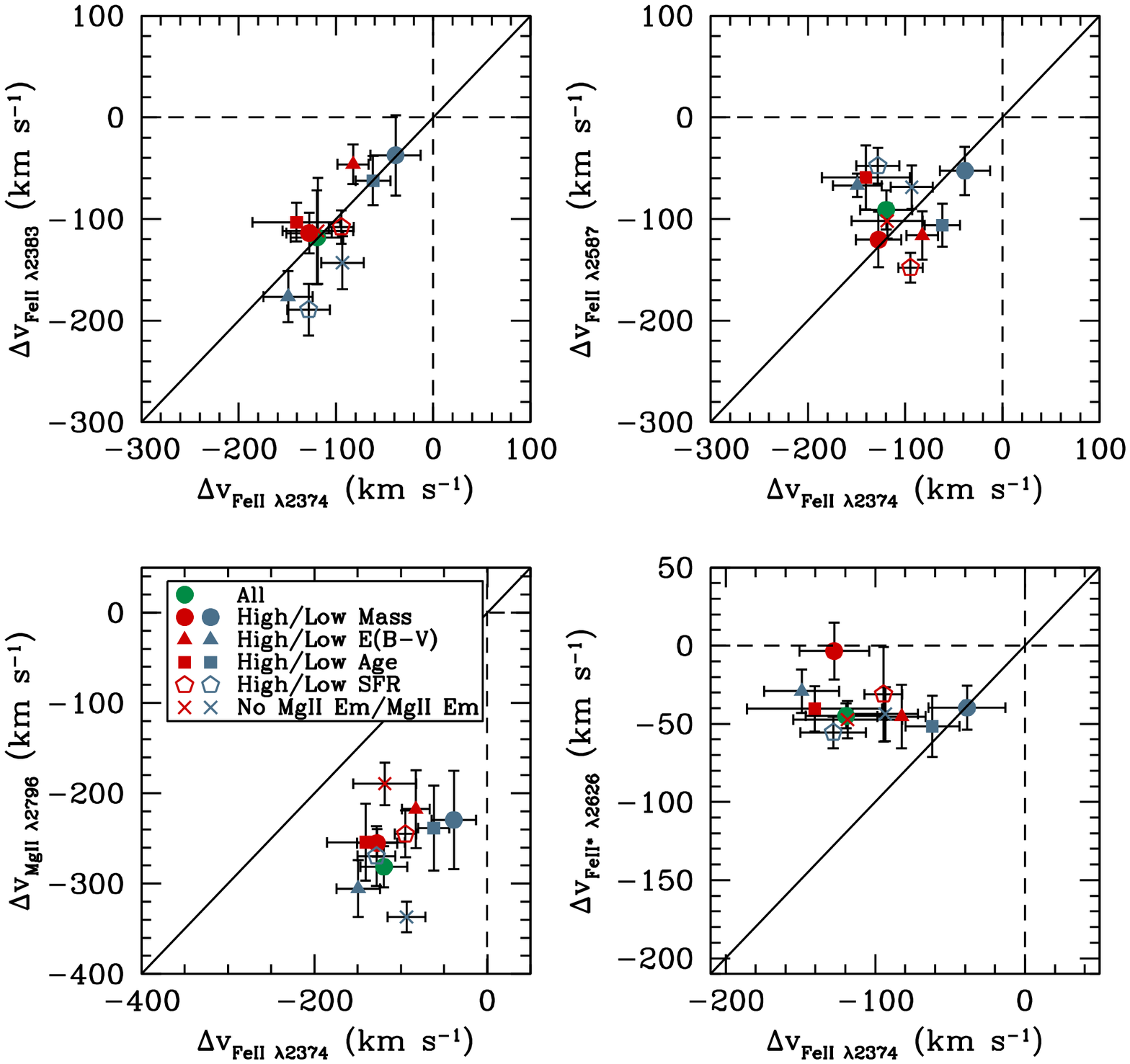}
\caption{Velocity comparisons for composite spectra. The velocities of \FeII\ $\lambda$2383 absorption (upper left),  \FeII\ $\lambda$2587 absorption (upper right),  \MgII\ $\lambda$2796 absorption (lower left) and  \FeII* $\lambda$2626 emission (lower right) are each plotted against the velocity of \FeII\ $\lambda$2374 absorption.  Compare Figure  \ref{fig:ind_velcomp}, for the same measurements using individual spectra.}
\label{fig:compvels}
\end{figure*}

The most striking feature of the composite spectra shown in Figure \ref{fig:allcomposites} is the difference in \MgII\ profile, particularly for the high and low mass composites. The lowest mass galaxies have strong \MgII\ emission and weak absorption, while the most massive galaxies have stronger absorption and no apparent emission.  Galaxies with young ages, low dust content, and (marginally) lower SFRs also show stronger \MgII\ emission, although the differences are not as striking as in the composites based on stellar mass.  Similar trends were seen by \citet{wcp+09}, who saw increased \MgII\ emission in composite spectra of lower mass galaxies and galaxies with lower SFRs (although \citealt{wcp+09} removed objects with excess \MgII\ emission from their sample, additional emission became apparent when the symmetric, zero-velocity absorption component was removed from the \MgII\ profile; we do not attempt such decomposition here).  The composite spectra also show more subtle differences in the strengths of the other absorption and emission features.  We discuss these differences below.

\subsection{Relative Velocities and Equivalent Widths of Absorption and Emission Lines in Composite Spectra}
\label{sec:velw_compspec}
An inspection of the absorption line measurements in Table \ref{tab:abs} shows that the most consistent differences between the composite spectra  arise when galaxies are divided based on stellar mass.  All of the lines are blueshifted to larger velocities in the high mass spectrum relative to the low mass, and most of the lines have higher equivalent widths in the high mass galaxies as well.  This larger blueshift suggests that more massive galaxies drive faster outflows, although we caution that the velocity centroid of the line is a poor proxy for outflow velocity because it may be influenced by absorption from the interstellar medium of the galaxy or by underlying emission.  We discuss the question of outflow velocity in more detail in Section \ref{sec:maxvels}.

\subsubsection{Relative Velocities of Absorption and Emission Lines}
\label{sec:vel_compspec}
In Figure \ref{fig:compvels} we compare the velocity centroids of several of the lines in the various composite spectra.  This figure should be compared with Figure \ref{fig:ind_velcomp}, in which we made the same comparisons for the individual spectra and found some evidence for the presence of underlying emission (in $\sim1/3$ of the sample) in the form of  blueshifted \FeII\ $\lambda 2383$ velocities relative to \FeII\ $\lambda 2374$ and \FeII\ $\lambda 2587$; recall that the latter two lines are coupled to fine structure transitions and therefore may be less affected by emission filling.  This signature blueshift of emission filling is less apparent in the composites.  We see generally good agreement between all three of the \FeII\ absorption lines, though the scatter is significant, particularly in the comparison of FeII\ $\lambda 2374$ and FeII\ $\lambda 2587$.  We conclude that although underlying emission may affect the velocity centroids of some \FeII\ lines in some galaxies, the effect is not strong enough to be robustly detected in composite spectra.  This may be because the fraction of galaxies so affected is relatively small, or because the higher S/N of the composites in combination with the method of direct integration makes the measurements of the composites more sensitive to absorption at a wider range of velocities, thereby decreasing the effect of underlying emission near zero.

Measurements of the individual spectra showed that the velocity centroid of \MgII\ $\lambda 2796$ is more affected by underlying emission than any of the \FeII\ lines, and this is also true of the composites.  In contrast to the \FeII\ lines studied, \MgII\ is blueshifted with respect to \FeII\ $\lambda 2374$ in all of the composite spectra.  This shift is probably due to underlying emission.  The largest velocity difference is observed in the composites of galaxies with and without \MgII\ emission: the average blueshift of \MgII\ $\lambda 2796$ in galaxies with no detectable \MgII\ emission is $\Delta v = -189 \pm 24$ \kms, while the average shift of those with emission is $\Delta v = -337 \pm 17$ \kms.  We observed above that low mass galaxies show stronger \MgII\ emission, and that the emission is somewhat stronger in galaxies with young ages, low dust content, and lower SFRs (all shown in blue in Figure \ref{fig:compvels}) as well.  Figure \ref{fig:compvels} suggests that this increased emission in low mass galaxies is apparent in the velocity of \MgII\ absorption as well, since the blue points tend to be more displaced than the red points from the line of equal velocities.

Finally, we compare the velocity centroid of the \FeII* $\lambda 2626$ emission line with that of \FeII\ $\lambda 2374$ in the composites (we note that the velocity of the \FeII* $\lambda 2626$ line is not affected by the nearby \FeII* $\lambda 2632$ line because the two lines are well-separated, as can be seen from Figure \ref{fig:allcomposites}). The individual spectra showed that the velocity of \FeII* $\lambda 2626$ scattered around zero, with a marginal tendency for greater blueshifts; in the composities we see consistently blueshifted velocity centroids of about $-50$ \kms.  As an additional check on this line,  we measure the velocity offset of \FeII* $\lambda 2626$ in the composite spectrum of the 51 galaxies with \OII\ redshifts and find it to be consistent with the other composites at $-43$ \kms.  The difference between the individual spectra and the composites may again be due to greater sensitivity to weak blueshifted emission in the composites; see the discussion of the fine structure emission lines in Section \ref{sec:emission} below. In any case, as with the individual spectra, \FeII* $\lambda 2626$ is on average observed at velocities closer to systemic than any of the \FeII\ absorption lines.

\subsubsection{Equivalent Widths of \FeII\ Absorption Lines}
\label{sec:w_compspec}
\begin{figure}[htbp]
\plotone{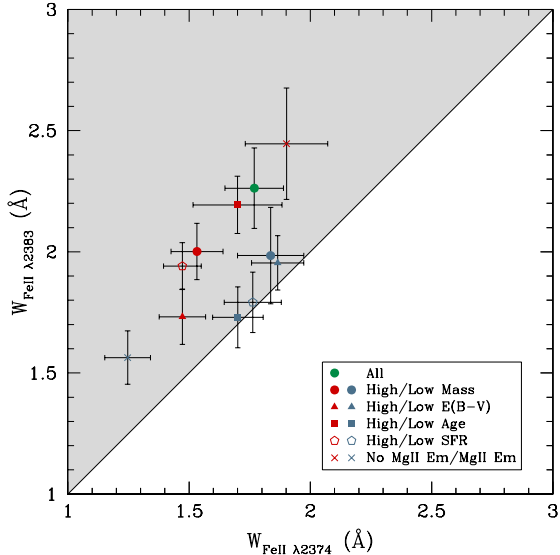}
\caption{The equivalent widths of the  \FeII\ $\lambda 2374$ and  \FeII\ $\lambda 2383$ absorption lines in the composite spectra, plotted with symbols given in the key at lower right.  The shaded region brackets the optically thin and optically thick line ratios (because the two lines have such different oscillator strengths, the upper limit of  $W_{\rm Fe II \lambda 2383}/W_{\rm Fe II  \lambda 2374} = 10.3$ does not fall within the region plotted). Compare Figure  \ref{fig:ind_ewcomp}, for the same measurements using individual spectra.}
\label{fig:comp_ew}
\end{figure}

Next we compare the equivalent widths of the  \FeII\ $\lambda 2374$ and  \FeII\ $\lambda 2383$ absorption lines in the composites, as we did for the individual spectra in Section \ref{sec:velw_indspec} and Figure \ref{fig:ind_ewcomp}. In the case of the individual spectra, we saw that, with the exception of four low-mass galaxies, all of the objects in the sample have an equivalent width ratio consistent with the range expected based on the variation of optical depth, $1 \leq W_{\rm Fe II \lambda 2383}/W_{\rm Fe II  \lambda 2374} \leq 10.3$.  The ratio of equivalent width for all of the composites falls within this range as well, as shown in Figure \ref{fig:comp_ew}.  We do see systematic differences in the equivalent width ratio, however: low mass, young, and unreddened galaxies with low SFRs (blue points) have $W_{\rm Fe II \lambda 2383}/W_{\rm Fe II  \lambda 2374} \approx 1$, while older, dustier, more massive galaxies with higher SFRs (red points) have $W_{\rm Fe II \lambda 2383}/W_{\rm Fe II  \lambda 2374} > 1$. These systematic differences may represent real differences in optical depth between the two samples, although it may perhaps be surprising that the lower mass and less reddened samples would show higher optical depth than the dustier and more massive galaxies.  The alternative explanation is that underlying emission is more likely to decrease the equivalent width of \FeII\ $\lambda 2383$ relative to \FeII\ $\lambda 2374$ in the low mass sample. This second explanation is supported by the four low mass galaxies whose individual spectra require the presence of such underlying emission (see Figure  \ref{fig:ind_ewcomp}), and perhaps by the fact that \MgII\ emission is clearly stronger in low-mass galaxies. However, such underlying emission must have relatively little effect on the velocity centroids of the lines, since systematic differences between the red and blue points are less apparent in the velocity comparison of FeII\ $\lambda 2374$ and  \FeII\ $\lambda 2383$ shown in Figure \ref{fig:compvels} than in the equivalent width comparison in Figure  \ref{fig:comp_ew}. 

\subsubsection{Maximum Velocities of Absorption Lines}
\label{sec:maxvels}
We next examine the maximum blueshifted velocities of the absorption lines in composite spectra, a measurement we could not make for the individual objects due to limited S/N.  High resolution, high S/N spectra of lensed galaxies show that the outflow-related interstellar absorption lines extend to typical velocities of $-700$ to $-800$ \kms\ \citep{psa+00,qpss09,qsp+10,dds+09}, and similar maximum velocities are seen in composite spectra of galaxies at \ztwo\ \citep{ses+10}.  \citet{wcp+09} measured the velocity $v_{10\%}$ at which \MgII\ $\lambda2796$ absorption reaches 90\% of the continuum level in galaxies at $z=1.4$, and found it to be correlated with both mass and star formation rate.  Although measurement of the maximum wind velocity is dependent on S/N, it is less likely to be affected by the interstellar medium or by underlying emission than the velocity centroid measurements discussed above.

We define the maximum velocity as the velocity corresponding to the wavelength at which the absorption profile first meets the normalized continuum (i.e., has a value of one) on the blue side of the line.  Because this measurement is highly dependent on S/N, we use the 5000 Monte Carlo simulations described above, and take $v_{\rm max}$ to be the average of the velocity at which the line meets the continuum in the 5000 realizations, with $1\sigma$ error given by the standard deviation.  We measure $v_{\rm max}$ for the five strong \FeII\ absorption lines and for \MgII\ $\lambda 2796$ (we do not measure \MgII\ $\lambda 2803$ because the blue wing is affected by the \MgII\ $\lambda 2796$ line). The average value of  $v_{\rm max}$ for these six lines in the composite spectrum of all 96 galaxies is $v_{\rm max} = -730$ \kms, in good agreement with other measurements of galaxies at \ztwo.  As other authors have discussed (e.g.\ \citealt{ses+10}), such a velocity is likely to be higher than the escape velocity of the galaxy, and it is therefore likely that at least some of the gas escapes the halo.

\begin{figure*}[htbp]
\plotone{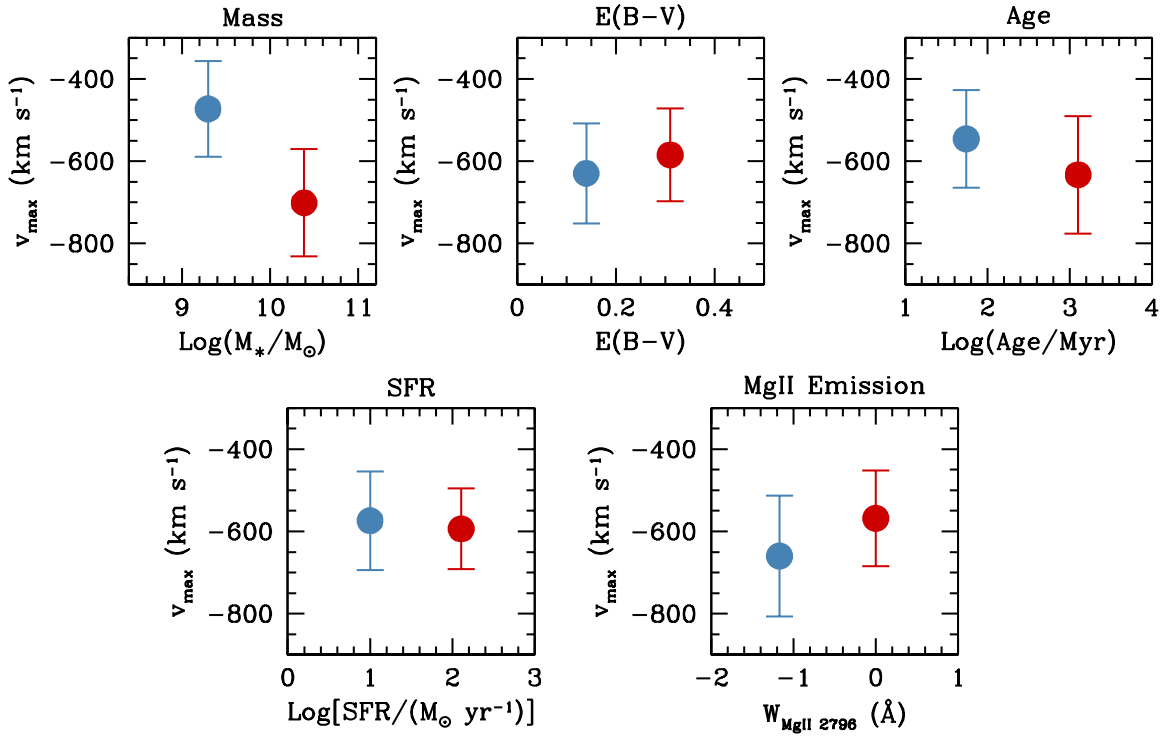}
\caption{The average maximum absorption velocity $v_{\rm max}$ in the composite spectra, measured as described in the text from the average of the five strong \FeII\ lines and \MgII\ $\lambda 2796$.}
\label{fig:termvels}
\end{figure*}

Figure \ref{fig:termvels} shows $v_{\rm max}$ for each of the composite spectra.  We have averaged the measurements of $v_{\rm max}$ for the six absorption lines, and plotted the result against the average value of the parameter on which the composite spectrum is based (e.g.\ the average stellar mass of the spectrum of low mass galaxies).  The uncertainties in this measurement are large, but the maximum outflow velocity shows little dependence on dust content, age, SFR or the presence of \MgII\ emission.  We do, however, see a weak dependence on stellar mass; this is the only parameter for which the maximum velocities of all six lines are higher in one composite relative to the other.  In the high mass composite, we find $v_{\rm max} = -701 \pm 131$ \kms, while in the low mass composite we measure $v_{\rm max} = -473 \pm 116$ \kms, for a difference of $\Delta v_{\rm max} = 228 \pm 124$ \kms.  Thus, we find that the maximum outflow velocity in massive galaxies is $\sim1.5$ times faster that that in low mass galaxies, with 1.8$\sigma$ significance.  For comparison, we find $\Delta v_{\rm max} =-45\pm117$ \kms\ for the $E(B-V)$ composites, $\Delta v_{\rm max} =20\pm110$ \kms\ for the SFR composites, $\Delta v_{\rm max} =87\pm132$ \kms\ for the age composites, and $\Delta v_{\rm max} =-92\pm132$ \kms\ for the composites based on \MgII\ emission.  We also note that all of the composites have similar S/N of 11--16 per pixel (see Table \ref{tab:sedcomps}); while the S/N of the high mass sample is somewhat higher than that of the low mass sample, this is also the case for several other pairs of composites, indicating that it is not likely to be the cause of the velocity difference observed between the high and low mass composites.

Using 1400 spectra of galaxies at $z\sim1.4$ from the DEEP2 survey, \citet{wcp+09} found that $v_{10\%} \propto M_{\star}^{0.17}$ and $v_{10\%} \propto {\rm SFR}^{0.3}$.  We clearly do not see the same dependence on star formation rate, as our measured values of $v_{\rm max}$ are nearly identical in the high and low SFR composites. Given the mean SFR values for the two composites (10 and 128 \msunyr), we would expect the high SFR sample to have a velocity 2.1 times higher that that of the low SFR sample, a result inconsistent with the observed $\Delta v_{\rm high\, SFR}/\Delta v_{\rm low\, SFR}=1.0\pm0.3$ at the level of 3.7$\sigma$ (though given the systematic uncertainties on the SFR discussed above, it would be desirable to repeat this measurement with an independent estimate of the SFR).  Similarly, in a sample of galaxies with \MgII\ and \FeII\ absorption at $z\sim1$, \citet{ksm+12} find only weak ($\sim1\sigma$) evidence for a trend between outflow velocity and star formation rate.  Some insight into the lack of correlation may be found by considering galaxies in the local universe.  Using observations of \ion{Na}{1} absorption in a sample of nearby galaxies spanning a range of four decades in SFR, \citet{m05} found a scaling of $v \propto {\rm SFR}^{0.35}$, while \citet{rvs05} suggest that the trend of increasing velocity with SFR flattens for SFRs above $\sim10$ \msunyr. If this is the case, high redshift samples probing a limited range in SFR may be unlikely to detect a relationship with velocity. We note, however, that the \citet{wcp+09} sample does not span a broad range in SFR; their low SFR bin contains galaxies with ${\rm SFR} < 14$ \msunyr, while their high SFR bins consists of galaxies with ${\rm SFR} > 28$ \msunyr.

Turning to our observed mass dependence, we find that it is entirely consistent with the results of \citet{wcp+09}, however; for the mean stellar masses of the upper and lower mass bins, a dependence of $v_{\rm max} \propto M_{\star}^{0.17}$ would result in the massive galaxies having a maximum velocity 1.5 times larger than the low mass galaxies, exactly the difference between our two measured values.  Given the uncertainties in our measurements of $v_{\rm max}$, however, a variety of mass dependencies would be consistent with our observations.
		 
\subsection{\FeII* and \MgII\ Emission}
\label{sec:emission}

\begin{figure*}[htbp]
\plotone{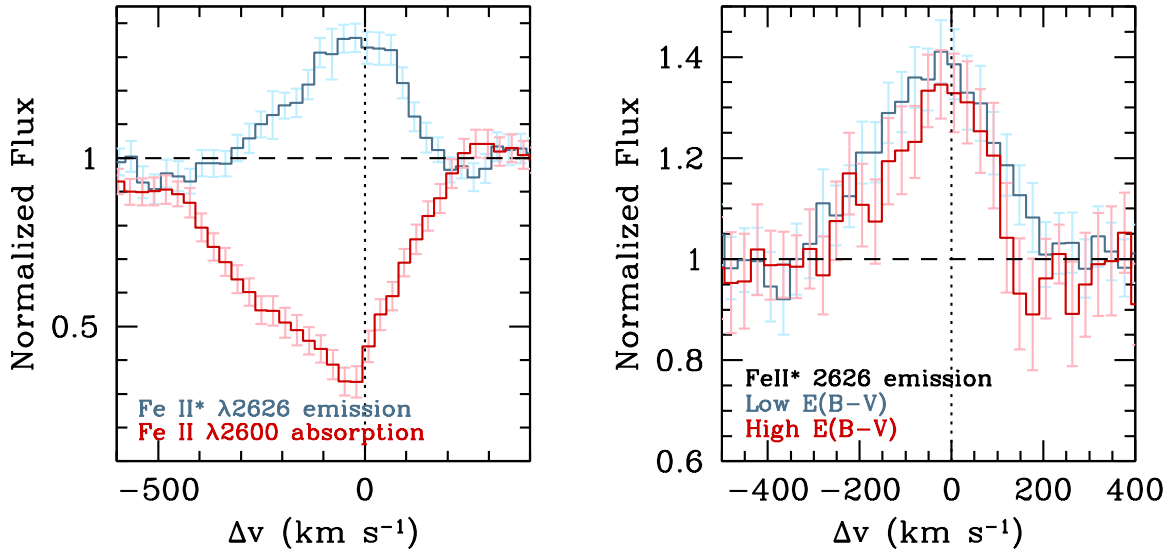}
\caption{{\it Left:} Velocity profiles for \FeII\ $\lambda$2600 absorption (red) and \FeII * $\lambda$2626 emission (blue), for the composite spectrum of all 96 galaxies.  The  \FeII\ $\lambda$2600 absorption profile does not meet the continuum on the blue side of the line because of adjacent  \FeII\ $\lambda$2587 absorption. {\it Right:} The  \FeII * $\lambda$2626 emission line profile in the high $E(B-V)$ (red) and low $E(B-V)$ (blue) composite spectra.}
\label{fig:emabs_vel}
\end{figure*}

The remainder of this paper is devoted to the emission features in the composite spectra. As listed in Table \ref{tab:lines} and shown in Figures \ref{fig:all_fullspec} and \ref{fig:allcomposites}, we detect emission from five \FeII* fine structure lines, and see significant emission in the \MgII\ $\lambda\lambda$ 2796, 2803 doublet for about $1/3$ of the galaxies in the sample.  \MgII\ emission is more likely to appear in galaxies with blue UV slopes and in lower mass galaxies.  Galaxies with \MgII\ emission also have stronger \FeII* emission; this can be seen most clearly in the composite spectra of galaxies with and without \MgII\ emission shown in the bottom panels of Figure  \ref{fig:allcomposites} and via the measured equivalent widths of the \FeII* lines in these spectra given in Table \ref{tab:em}.  Given the anti-correlation between \MgII\ emission and stellar mass, it is then unsurprising that low mass and low SFR galaxies also tend to have stronger \FeII* emission compared to their higher mass, higher SFR counterparts. 

\FeII* emission is generally not seen in local star-forming and starburst galaxies \citep{lthc11}. These lines have also not been seen in samples of galaxies at redshifts similar to the current sample, due to spectral coverage in the case of \citet{wcp+09} and probably due to S/N limitations in the sample of \citet{rwk+10}.  The situation is similar to that of the \SiII* fine structure lines in the far-UV, which are seen in galaxies at $z\sim2$--4 \citep{ssp+03,eps+10,jse12} but not in local starbursts \citep{smc+06}.

\citet{rpm+11} detect both \FeII* and \MgII\ emission in a starburst galaxy at $z=0.69$, and propose that both are generated by photon scattering in outflowing gas.  They also show that the \MgII\ emission is significantly spatially extended compared to the stellar continuum. This model may explain the absence of fine structure emission in local galaxies, if the slits used to observe such objects cover only the central regions of the galaxies and capture little emission from an extended outflow \citep{gvs+11}.  \citet{pkr11} use radiative transfer models to predict the strength of \FeII\ and \MgII\ absorption and emission for a variety of outflow models; we compare our results to these models below. 

As a first test of the origin of the \FeII* emission, we compare the velocity structure of the strongest line, \FeII* $\lambda 2626$, with that of the absorption line from which the emission is expected to arise, \FeII\ $\lambda 2600$, using the composite spectrum of all 96 galaxies. The comparison is shown in the left panel of Figure \ref{fig:emabs_vel}, with \FeII* $\lambda 2626$ emission in blue and \FeII\ $\lambda 2600$ absorption in red.  The absorption line shows the clear asymmetric profile of galactic outflows, with a blueshifted centroid and tail extending to large velocities.  \FeII* $\lambda 2626$ is a nearly perfect mirror of the absorption profile in emission, suggesting that the two lines do indeed arise from the same gas. The \FeII* $\lambda 2626$ profile also indicates the presence of dust; since the \FeII* lines are optically thin (we see no absorption at their wavelengths), \FeII* emission should reach us from both sides of the outflow, rather than only the near side seen in absorption. The fact that we see more emission from the blueshifted side than the redshifted side is probably due to extinction, as photons from the far side of the galaxy must pass through more dust in order to reach us \citep{pkr11}. As a test of this scenario, we also compare the \FeII* $\lambda 2626$ profiles in the high and low $E(B-V)$ composite spectra, in the right panel of Figure \ref{fig:emabs_vel}. If the asymmetric line profile is due to extinction, we would expect greater asymmetry in the composite spectrum of the dustier galaxies. The lower S/N of the extinction-based composites makes a robust comparison difficult, but Figure \ref{fig:emabs_vel} does show that the red wing of the line extends to somewhat higher velocities in the low $E(B-V)$ composite.  A comparison of the equivalent widths of the blue and red sides of the lines also offers support to the model in which the line asymmetry is due to extinction; the red flux is $\sim55$\% of the blue flux in the high $E(B-V)$ composite, and  $\sim65$\% of the blue flux in the low $E(B-V)$ composite.

In this section we examine the origins of \FeII* and \MgII\ emission in more detail, through photoionization modeling (Section \ref{sec:cloudy}), a study of the spatial extent of the emission (Section \ref{sec:spatialextent}), and a comparison of the relative strengths of the \FeII\ absorption lines and their associated \FeII* emission (Section \ref{sec:emabs}).

\subsubsection{Photoionization Modeling}
\label{sec:cloudy}

\begin{figure*}[htbp]
\plotone{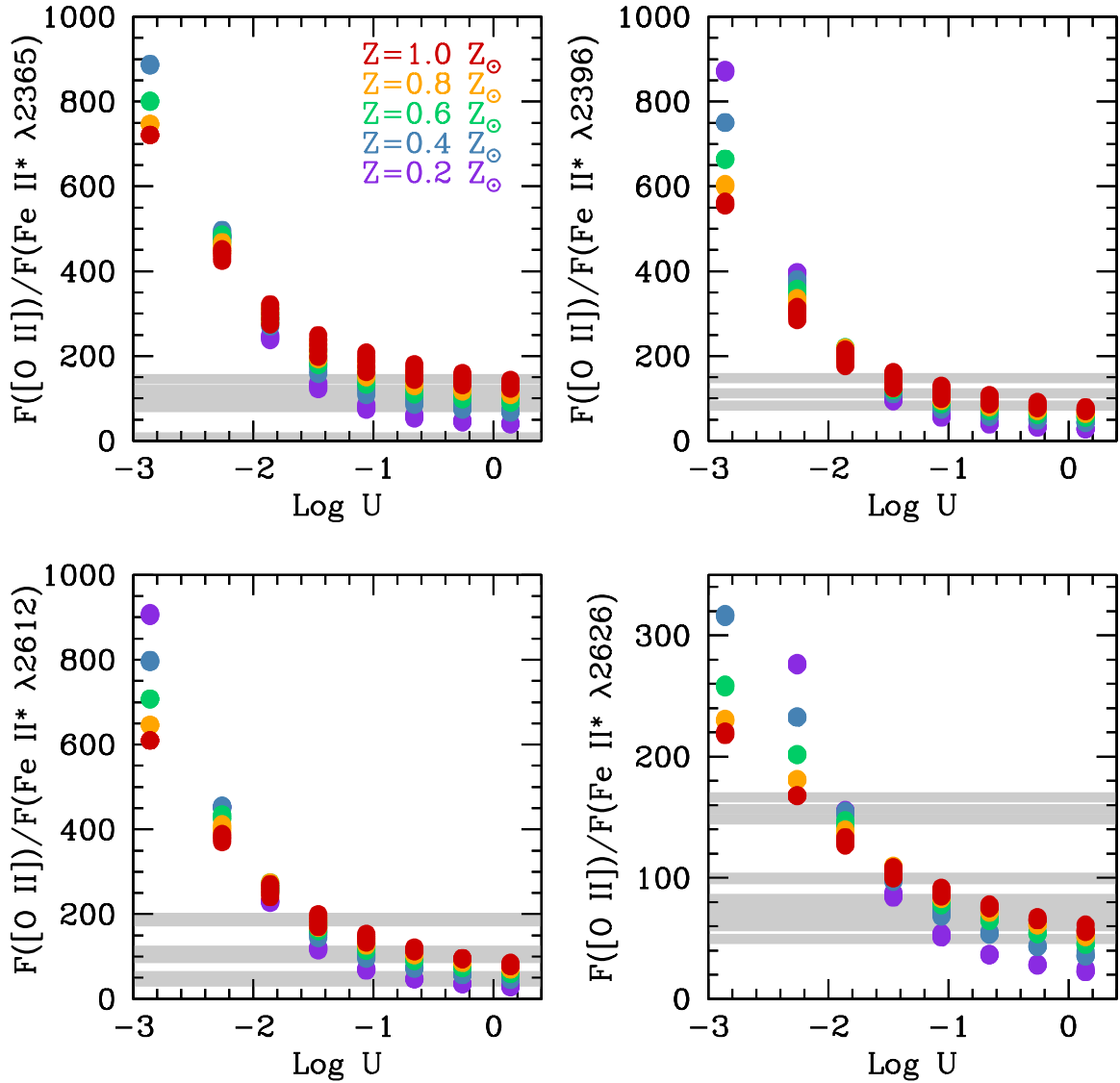}
\caption{Colored points show the predicted ratios of the luminosities of the \FeII * fine structure emission lines to \OII\ $\lambda\lambda$ 3726, 3729 emission, plotted against ionization parameter.  The O/Fe ratio is the same for all models shown, corresponding to Cloudy's default \HII\ region abundance ratio. The points are color-coded by metallicity according to the key in the upper left panel.  Grey horizontal lines show the measured line ratios in individual spectra.}
\label{fig:cloudy}
\end{figure*}

In this subsection we use photoionization models to study the origins of \FeII* and \MgII\ emission.  While previous work \citep{rpm+11,pkr11} and the kinematics of the \FeII* $\lambda 2626$ line suggest that \FeII* fine structure emission arises in the outflow, other sources are also possible. In order to assess the possibility that the \FeII* emission originates in \HII\ regions, we have constructed a suite of photoionization models which we use to examine the strength of \FeII* emission as a function of metallicity and ionization parameter.  The models were constructed with version 08.00 of Cloudy, last described by \citet{cloudy}. We employ an input ionizing spectrum from Starburst99 \citep{sb99}, assuming continuous star formation and a metallicity of $Z = 0.008$, or $\sim 0.6 Z_{\odot}$. We assume an electron density $n_e=100$ cm$^{-3}$, consistent with the density inferred from the average value of the density-sensitive \OII $\lambda 3929$/\OII $\lambda 3926$ line ratio. (We find $\langle$\OII $\lambda 3929$/\OII $\lambda 3926\rangle = 1.4 \pm 0.1$, implying densities between the low density limit and $n_e \sim  140$ cm$^{-3}$.)  We consider gas phase metallicities of 0.2, 0.4, 0.6, 0.8, and 1.0 $Z_{\odot}$.   We use Cloudy's default \HII\ region abundance set, and in addition consider models with the Fe abundance reduced by factors of 0.8, 0.6, 0.4, and 0.2 for each of the metallicities adopted.  Each model is characterized by an ionization parameter $U$, which is defined as the ratio of ionizing photon density to total hydrogen density,
\begin{equation}
U \equiv \frac{Q}{4\pi r_0^2 \, n \, c},
\end{equation}
where $Q$ is the rate of ionizing photons (number s$^{-1}$, given by the Starburst99 model used to construct the input ionizing spectrum), $r_0$ is the distance from the ionizing source to the inner edge of the cloud, and $n$ is the hydrogen density. The models assume a spherical \HII\ region, but the geometry is effectively plane parallel if the thickness of the cloud is less than $0.1r_0$; this is the case for all of our models except those with the highest values of the ionization parameter considered, $\log U \sim 0$. The models use Cloudy's default stopping criterion, in all cases stopping when the temperature falls below 4000 K.

Results of the photoionization modeling are shown in Figure \ref{fig:cloudy}. We plot the ratio of each of the \FeII* lines to \OII\ $\lambda\lambda$3626, 3729 against the logarithm of the dimensionless ionization parameter $U$, which represents the ratio of ionizing photons to particles. (It is more standard to measure the strength of a line of interest against a recombination line such as \Hb, but \OII\ is the only strong nebular emission line for which we have measurements.)  The \OII/\FeII* ratio shows little metallicity dependence but decreases strongly with increasing ionization parameter, driven largely by the decrease in \OII\ (and corresponding increase in \OIII) as ionization parameter increases. Grey horizontal lines show the measured ratios for individual spectra in our sample. 

\begin{figure*}[htbp]
\plotone{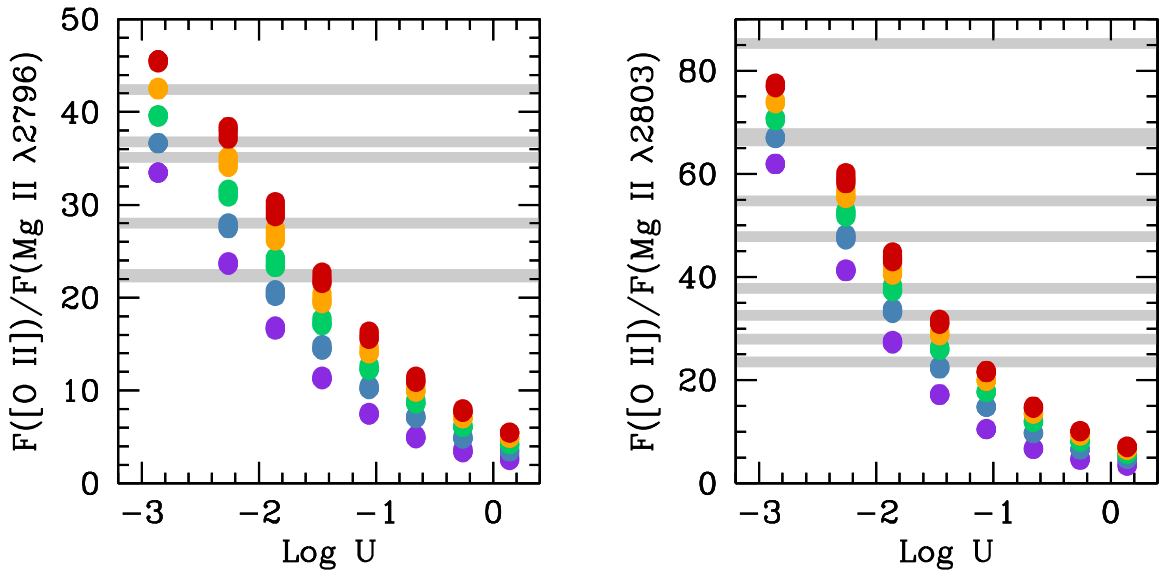}
\caption{Colored points show the predicted ratios of \MgII\ emission to \OII\ $\lambda\lambda$ 3726, 3729 emission, plotted against ionization parameter.  \MgII\ $\lambda2796$ is shown in the left panel and  \MgII\ $\lambda2803$ in the right panel. The points are color-coded by metallicity as in Figure \ref{fig:cloudy}.  Grey horizontal lines show the measured line ratios in individual spectra.}
\label{fig:cloudy_mg2}
\end{figure*}

It is clear that, in order to match the observed ratios, a very high ionization parameter of $\log U \gtrsim -2$ is required. We do not have ionization parameter measurements for the galaxies in our sample, but measurements of lensed galaxies at \ztwo\ suggest that typical star-forming galaxies at these redshifts have $-2.8 \lesssim \log U \lesssim -2.3$ \citep{hsk+09}, higher than values observed in most local galaxies but lower than needed to produced our observed \OII/\FeII* ratios.  We also note two additional factors. First, the measured line ratios are not corrected for dust extinction. This is unlikely to be a large effect, given the relatively narrow wavelength separation, but the effect of dust would decrease the \FeII* lines relative to \OII, making the observed \OII/\FeII* ratios higher than the intrinsic ratios and further increasing the ionization parameter needed to match the data.  Second, all of the models shown assume Cloudy's default \HII\ region O/Fe ratio, which may not be appropriate given that an underabundance of the Fe-peak elements relative to $\alpha$-capture elements (such as O) has been observed in lensed galaxies at high redshifts \citep{prs+02, dds+09}.  We have tested the effect of a decreased abundance of Fe relative to O in the Cloudy models; the effect is to weaken the \FeII* emission, increase the  \OII/\FeII* ratio, and thereby increase the ionization parameter needed to match the observations.  In other words, a decreased abundance of Fe relative to O shifts all of the colored points in Figure \ref{fig:cloudy} upwards. 

In summary, it is unlikely that that the galaxies in our sample would have ionization parameters high enough to produce the required \OII/\FeII* ratios, and it is  therefore also unlikely that the observed \FeII* emission originates in the \HII\ regions of the galaxies.  

Next we examine the question of \MgII. In contrast to \FeII* (and resonant \FeII) emission, emission from \MgII\ is observed in local \HII\ regions \citep{kbc+93}, so it is important to assess the likely sources of this emission in high redshift galaxies. In Figure \ref{fig:cloudy_mg2} we plot the predicted ratios of \OII\ to \MgII\ $\lambda2796$ and $\lambda2803$ emission against the ionization parameter, color-coding by metallicity as in Figure \ref{fig:cloudy}.  The observed ratios are shown by the grey horizontal lines; unlike the \OII/\FeII* ratios, they are fully consistent with the expected ionization parameters of $\log U \sim -3$ to $-2$.  The photoionization modeling therefore indicates that, in contrast to the \FeII* emission, much of the \MgII\ emission may arise in \HII\ regions. Because \MgII\ is a resonance line, however, photons produced in \HII\ regions then scatter in outflowing gas, producing the observed blueshifted absorption and redshifted emission profiles.   

We have suggested above (Section \ref{sec:individual}) that \MgII\ emission in star-forming galaxies at $z\sim1$--2 is similar to \lya\ emission at \ztwo--3, in that it exhibits a complex profile of combined emission and absorption; we have also seen that \MgII\ emission is more common in low mass and less dusty galaxies.  The photoionization modeling supports this connection, indicating that \MgII\ emission, like that of \lya, arises from the resonant scattering of photons produced in \HII\ regions.  Thus the production of \MgII\ photons in \HII\ regions is likely to account for the difference between the profiles of \MgII\ and that of \FeII\ lines not coupled to fine structure transitions such as \FeII\ $\lambda2383$. For example, the spectra of Q1623-BM95 and Q1700-BM154, shown in Figures \ref{fig:ind_fespecs} and \ref{fig:mg2profiles}, show weak to moderate \FeII\ $\lambda2383$ absorption with no associated emission, while their \MgII\ profiles show strong redshifted emission accompanied by weak blueshifted absorption. Under the assumption that emission is produced by the scattering of continuum photons produced in the outflow, one would expect to see similar emission in the \MgII\ and \FeII\ $\lambda2383$ lines; instead we see much stronger emission in the \MgII\ transitions, accounted for by the production of \MgII\ photons in \HII\ regions and their subsequent resonant scattering.  A comparison of the predicted strengths of \MgII\ $\lambda2796$ and \FeII\ $\lambda2383$ emission in the photoionization models supports this conclusion; for ionization parameters $\log U \sim -3$ to $-2$, emission from MgII\ $\lambda2796$ is $\sim7$ times stronger than 
\FeII\ $\lambda2383$ emission. We therefore expect scattered emission from \HII\ regions to be much more significant for \MgII\ than for \FeII\ $\lambda2383$.

More detailed photoionization modeling in the future may address the question of whether the \OII\ and \MgII\ emission arise from the same parts of the \HII\ regions; according to \citet{pkr11}, \MgII\ emission from \HII\ regions arises primarily from recombinations in the outer layers. An additional question which cannot be addressed by our current models and data is that of additional sources of \OII\ emission. \OII\ emission from diffused ionized gas is seen in local galaxies (e.g.\ \citealt{martin97}), and is likely present in higher redshift galaxies as well. However, the Cloudy models do not provide predictions for such emission or absolute \OII\ luminosities with which to compare our measurements, and our observations do not have the spatial resolution necessary to address this question (though see Section \ref{sec:spatialextent} below, where we show that the \OII\ emission has the same spatial distribution as the stellar continuum, indicating that most of the emission does not arise in an extended outflow).  Our observed \OII/\MgII\ ratios are consistent with emission from both ions arising in \HII\ regions, but other sources of \OII\ emission cannot be ruled out. Future, more detailed modeling and observations with higher sensitivity and spatial resolution will address this question.

To summarize the results of this subsection, the photoionization modeling indicates different mechanisms for the production of \FeII* and \MgII\ emission. Although both ultimately arise in the outflow, \FeII* emission arises from the re-emission of continuum photons absorbed in the \FeII\ resonance transitions, while the resonant scattering of photons originally produced in \HII\ regions accounts for much of the \MgII\ emission.

\subsubsection{The Spatial Extent of \FeII* and \MgII\ Emission}
\label{sec:spatialextent}

\begin{figure*}[htbp]
\epsscale{0.9}
\plotone{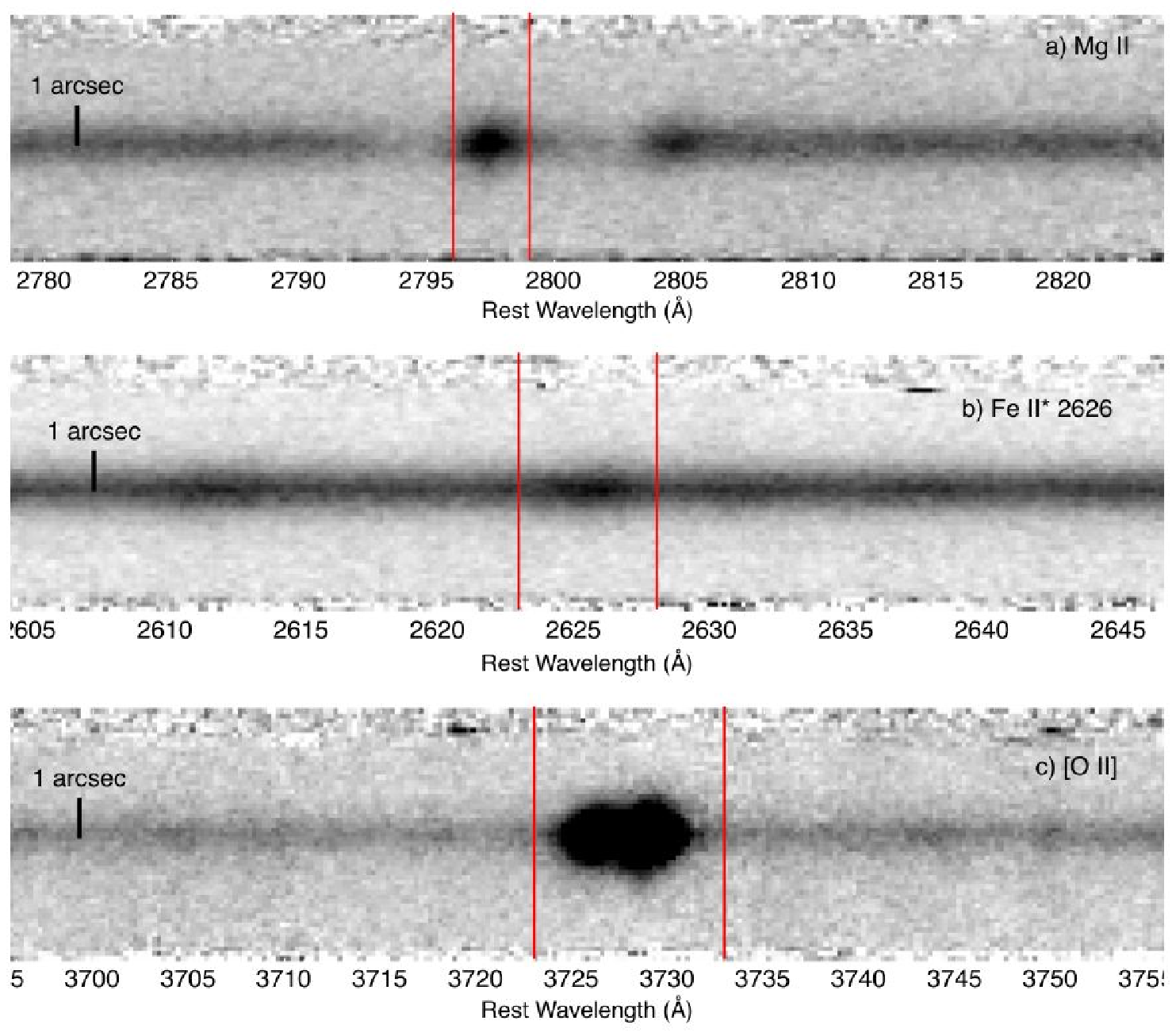}
\caption{Two-dimensional composite spectra of \MgII\ emission (top), \FeII* $\lambda 2626$ emission (middle), and \OII\ emission (bottom). The red vertical lines show the wavelength range over which the spatial profiles are measured.}
\label{fig:2dspecs}
\end{figure*}

\begin{figure*}[htbp]
\plotone{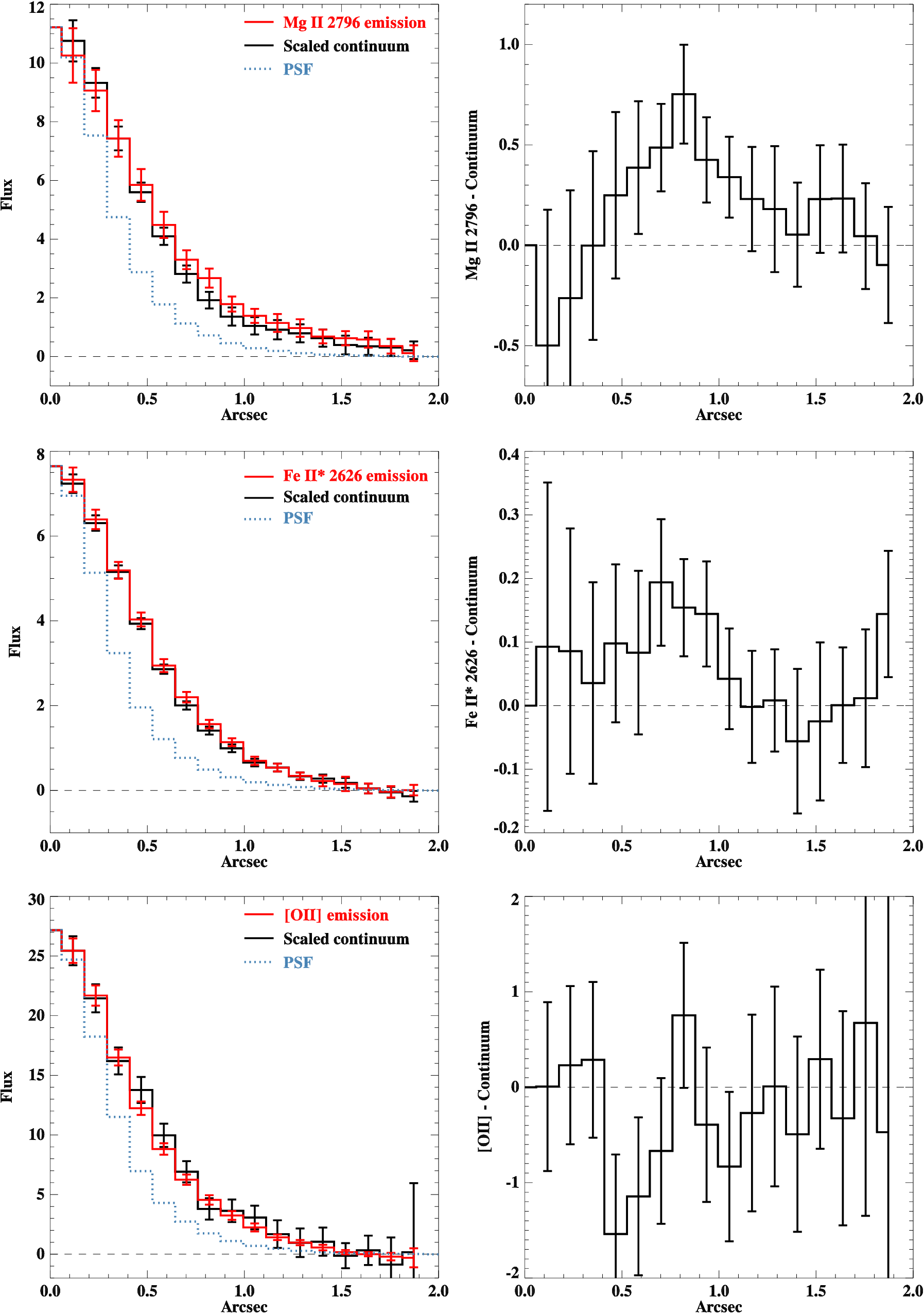}
\caption{The spatial profiles of \MgII\  $\lambda$2796 (top), \FeII * $\lambda$2626 (middle), and \OII\ $\lambda\lambda3727$, 3729 emission (bottom) compared to the continuum.  In the left panels the emission lines are shown in red and the continuum in black.  The right panels show the continuum subtracted from the emission.  Note the differing $y$-axis scales.}
\label{fig:spatialprofiles}
\end{figure*}

If the \FeII* emission originates in a large-scale galactic outflow, it may have a more extended spatial distribution than the stellar continuum; similarly, resonant scattering of \MgII\ photons in outflowing gas may produce spatially extended \MgII\ emission \citep{pkr11}.   \citet{rpm+11} study \FeII* and \MgII\ emission in a galaxy at $z=0.69$, finding that \MgII\ emission is broader than the stellar continuum and extends  to distances of $\gtrsim 7$ kpc.  Likewise, \citet{sbs+11} use deep narrowband images to show that diffuse \lya\ emission surrounds galaxies at \ztwo\ to distances of $\sim 80$ kpc;  the \lya\ transition is similar to \MgII\ in its radiative transfer properties.

In order to assess the spatial extent of \MgII\ and \FeII* emission in our sample we create two-dimensional composite spectra and compare the spatial extent of the line emission relative to the stellar continuum.  We also construct a 2D composite  \OII\ spectrum for comparison. The 2D composite spectra are created by cutting out a 100 \AA\ region around the line of interest in each individual spectrum. These cutouts are also centered spatially (but without spatial resampling) on the object, using the object positions stored in the image header by the DEEP2 reduction pipeline.  The cutouts are then resampled to a common dispersion of 20 \kms\ per pixel using linear interpolation, and the appropriate 2D wavelength solution is applied to each cutout.  Because the original two-dimensional spectra are rectified and because we consider only a short range in wavelength, the object traces are not curved and the wavelength does not depend on spatial location along the slit.  We then measure the centroid of the object position in each spectrum (since the positions in the header are not exact), and use these measurements to construct spatial offsets used to combine the individual spectra.  Further details of the combination depend on the line considered and are described below.

The \MgII\ composite includes only the 33 galaxies with \MgII\ emission,  determined as described in Section \ref{sec:mg2uvcolor}.  Because the peak velocity of \MgII\ emission varies significantly in individual spectra, we shift each 2D spectrum both spectrally and spatially, centering on the peak of the emission, and combine the individual spectra using a S/N-weighted average.  We measure the extent of \FeII* emission using the strongest of the \FeII* lines, \FeII* $\lambda 2626$.  We combine the full sample (95 of the 96 galaxies in the sample have spectral coverage of the  \FeII* $\lambda 2626$ line), aligning each spectrum spatially and again combining using a S/N-weighted average.  For the \OII\ composite, we combine the 51 galaxies with \OII\ emission, again aligning each spectrum spatially and combining using a S/N-weighted average.  In all three cases we then average the two sides of the composites in order to increase S/N.  We determine uncertainties by creating 500 artificial 2D composite spectra using bootstrap resampling: we measure the average and standard deviation of the line and continuum profiles in the 500 artificial spectra, and take the $1\sigma$ uncertainties in the spatial profiles to be the standard deviation of their measurements in the sample of artificial spectra.  The two-dimensional composite spectra for the three lines are shown in Figure \ref{fig:2dspecs}.

Spatial profiles of the line emission and stellar continuum are shown in Figure \ref{fig:spatialprofiles}. The top two panels show \MgII\ emission: at left, we compare the \MgII\ emission profile (red) with that of the stellar continuum (black), where the continuum has been scaled by a factor of 1.6 to have the same peak value as the \MgII\ emission.  The blue dotted line shows the point spread function, measured from the average spatial profiles of two stars appearing on the masks.  In the upper right panel we show the difference between the line and continuum emission.  The \MgII\ emission is slightly stronger at larger distances, reaching a peak excess of 40\% greater than the scaled continuum at a radius of 0.8 arcsec.

In the middle two panels we show the same comparison for the weaker \FeII* emission. In this case we scale the stellar continuum by a factor of 1.15 to match the peak \FeII* flux, and find that the \FeII* emission is also somewhat more extended than the continuum, though with a smaller excess; the difference between line and continuum peaks at $\sim0.7$ arcsec, with a 10\% excess in the line flux relative to the scaled continuum. Although this is a smaller excess than we see in the \MgII\ emission, we detect it with comparable significance because of the larger sample of galaxies used in the \FeII* composite. Because the \FeII* emission lines are stronger in galaxies with \MgII\ emission (see Section \ref{sec:emission}), a natural additional test would be to measure the excess \FeII* emission in galaxies with \MgII\ emission; unfortunately, however, the sample of \MgII\ emitters is too small to enable sufficient S/N to make this measurement.

The bottom two panels of Figure \ref{fig:spatialprofiles} show the comparison of the spatial profile of \OII\ emission relative to the continuum. \OII\ emission is expected to arise primarily in \HII\ regions, and because it is not a resonance line, its spatial distribution should trace that of the star-forming regions of the galaxy, with any differences with respect to the continuum emission arising from differing spatial distributions of the hot young stars producing the \HII\ regions and the older stars responsible for the continuum (or, at larger radii, the possible generation of \OII\ emission in the outflow). Figure \ref{fig:spatialprofiles} shows that the \OII\ emission indeed follows the spatial distribution of the continuum closely. There is no excess of line emission at $\sim1$ arcsec, as is observed for \MgII\ and \FeII* emission; instead the \OII\ emission is slightly weaker than the continuum at $\sim0.5$ arcsec, although the significance of this result is marginal. We conclude from this test of the spatial distribution of \OII\ emission that the observed excesses in \MgII\ and \FeII\ are not an artifact of the comparison of stronger line emission with weaker continuum.

We have presented evidence that both \MgII\ and \FeII* (but not \OII) emission are slightly more spatially extended than the underlying stellar continuum, with excess relative to the continuum detected at the $\sim1.5$--2$\sigma$ level.  This supports the conclusion that these emission lines arise in outflowing gas, though higher S/N measurements are clearly needed.  The S/N of the stacked spectra we use to make this measurement is only just high enough to detect a difference in the spatial distribution of lines and continuum, but not high enough to measure line emission to significantly larger radii than the continuum.  Detections of extended \lya\ emission to distances of $\sim 80$ kpc \citep{sbs+11} suggest that such emission could be present at much larger distances, but the weakness of \MgII\ and \FeII* emission relative to \lya\ mean that such detections at high redshift are likely out of the reach of current technology.

\subsubsection{Strengths of \FeII* Emission and \FeII\ Absorption Lines}
\label{sec:emabs}
We conclude our discussion of the origins of \FeII* emission with a comparison of the relative strengths of the \FeII* emission lines and their associated \FeII\ absorption.  As shown in Figure \ref{fig:energylevels}, each of the fine structure emission lines is coupled to an absorption line which provides the source of the emitted photons. The generic expectation for the model in which \FeII* emission is produced by photons absorbed in \FeII\ transitions in the outflow is that \FeII* emission will increase as \FeII\ absorption increases. \citet{pkr11} provide extended discussion of this issue, and use radiative transfer models to predict the relative strengths of absorption and emission lines for a variety of outflow models. For a spherically symmetric outflow in the absence of dust, conservation of photons requires that the total emission and absorption equivalent widths sum to zero; thus, in the absence of dust or particular geometric effects, we expect the total emission and absorption strengths to be approximately the same. Note also that the absorbed photons may be re-emitted as emission at their original wavelength rather than through one of the fine structure transitions, producing a P Cygni-like profile; whether or not this is likely to happen depends on the probabilities of the transitions involved and on the details of the galaxy, including the spatial and velocity distribution of outflowing gas, the presence of an interstellar medium at zero velocity, and the dust content.  

\begin{figure}[htbp]
\plotone{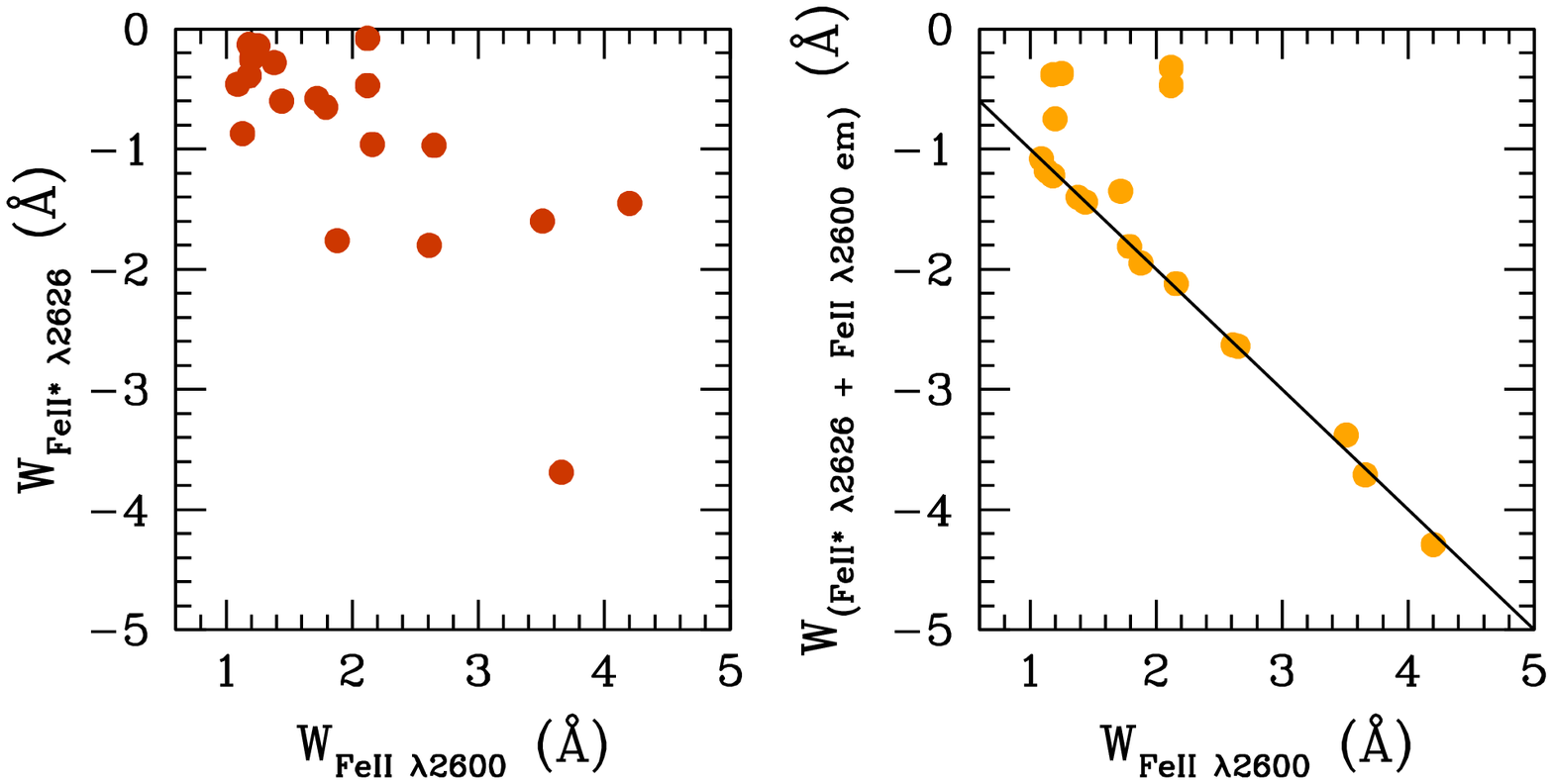}
\caption{Equivalent widths of \FeII\ $\lambda$2600 and \FeII * $\lambda$2626, as predicted by the models of \citet{pkr11}.  In both panels the $x$ axis shows the equivalent width of \FeII\ $\lambda$2600 absorption.  In the left panel, the $y$ axis shows \FeII * $\lambda$2626 emission, and in the right panel the $y$ axis shows the sum of the equivalent widths of \FeII * $\lambda$2626 emission and \FeII\ $\lambda$2600 emission.  The solid line in the right panel shows equal emission and absorption.  Most of the scatter in the relationship between  \FeII\ $\lambda$2600 absorption and  \FeII * $\lambda$2626 emission is due to the presence of \FeII\ $\lambda$2600 emission.  }
\label{fig:pkr}
\end{figure}

We show the expected  relationship between absorption and emission in Figure \ref{fig:pkr}, which shows the predicted equivalent widths of the  \FeII\ $\lambda$2600 and \FeII * $\lambda$2626 lines in all of the models considered by \citet{pkr11}.  The points represent a wide variety of outflow scenarios rather than a range  of galaxies within the same model; thus they should be taken simply as a generic prediction of how increasing one line affects the other. The left panel compares the strengths of \FeII\ $\lambda$2600 absorption and \FeII * $\lambda$2626 emission, and shows that, although there is considerable scatter, \FeII * $\lambda$2626 emission generally increases as  \FeII\ $\lambda$2600 absorption increases. The right panel shows that much of the scatter is due to the fraction of emission which escapes at 2600 \AA\ rather than at 2626 \AA; except in cases with significant extinction or particular outflow geometry, the sum of the total emission and absorption is approximately zero.  The outliers (points falling far from the solid line in the right panel of Figure \ref{fig:pkr}) represent either non-spherically symmetric models or models with significant dust content. These models show that either dust or geometry alone can cause large deviations from the simple conservation of photons.

\begin{figure*}[htbp]
\plotone{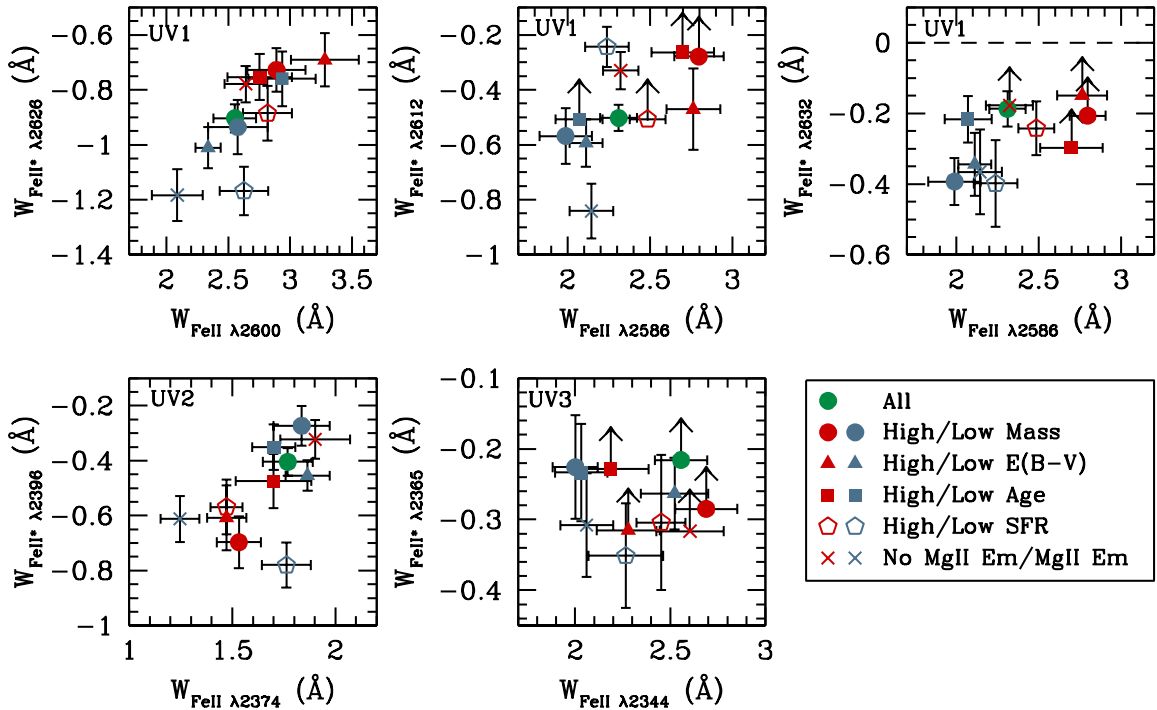}
\caption{The equivalent widths of \FeII* emission lines ($y$ axis in each panel) compared with the equivalent widths of the \FeII\ absorption lines from which the emission is assumed to arise ($x$ axis in each panel).  The points represent measurements of composite spectra, as shown in the legend at lower right.}
\label{fig:emabs_ew}
\end{figure*}

In Figure \ref{fig:emabs_ew} we plot the equivalent width of each of the \FeII* emission lines against its associated absorption line, for each of the composite spectra (note that because of the varying overlap of the samples used for the various composites, as discussed in Section \ref{sec:compspecs}, the points on these plots are not all independent).   We focus the following discussion primarily on the  \FeII\ $\lambda$2600 absorption and \FeII* $\lambda$2626 emission lines, since \FeII* $\lambda$2626 is the strongest of the \FeII* lines, but (with some exceptions noted below) the paired lines follow similar trends.
		
As already noted, \FeII* emission is stronger in less massive and less dusty galaxies and in galaxies with \MgII\ emission.  It is also immediately apparent that we see a trend of {\it decreasing} \FeII* $\lambda$2626 emission with increasing  \FeII\ $\lambda$2600 absorption, the opposite of what we expect based on the model described above.  We have not included \FeII\ $\lambda$2600 emission in this comparison, since it is not significantly ($>3\sigma$) detected in any of the composites; however, examination of Figure \ref{fig:allcomposites} shows some evidence for redshifted P Cygni  \FeII\ $\lambda$2600 emission in the low mass and low $E(B-V)$ composites.  These are the only spectra in which the emission is detected at $>2\sigma$ significance; we find $W_{\rm em}^{2600} = -0.21 \pm 0.08$ \AA\ in the low mass composite, and $W_{\rm em}^{2600} = -0.16 \pm 0.08$ \AA\ in the low $E(B-V)$ composite.  These two composites already show relatively weak \FeII\ $\lambda$2600 absorption and strong \FeII* $\lambda$2626 emission, so including emission at 2600 \AA\ simply increases the imbalance between total emission and absorption in these samples.  More generally, \FeII\ $\lambda$2600 emission decreases as the optical depth of the transition increases, and the \citet{pkr11} models show that little to no emission is seen when an ISM component is present in the galaxy.   Our observations therefore suggest higher optical depth in higher mass and dustier galaxies, possibly due to a more significant interstellar medium. \citet{wcp+09} also find that symmetric \MgII\ absorption centered at zero velocity is significantly stronger in more massive galaxies, and \citet{ses+10} find more absorption at $v\sim0$ \kms\ in galaxies with higher baryonic (gas + stellar) and dynamical mass at $z\sim2$.  A more significant ISM in more massive galaxies does not account for the observed decrease in \FeII* emission with increasing \FeII\ absorption, however, since the presence of an ISM increases both the absorption and total emission, while shifting the emission from the resonance lines to the fine structure transitions.

The fact that the total absorption is stronger than the total emission in all spectra may be attributable to dust; the \citet{pkr11} models indicate that the effect of dust is to suppress the emission with relatively little effect on the absorption profile.  As noted above, the \FeII* $\lambda$2626 emission profile in the composite spectrum of the full sample shows evidence for extinction in the lack of redshifted emission relative to blueshifted emission.  Furthermore, extinction may at least in part explain the observed trend in $W_{\rm abs}$ vs.\ $W_{\rm em}$, if dust attenuates a larger fraction of the emission in massive galaxies.  This does not at first appear to be a satisfactory explanation, since the average value of $E(B-V)$ is slightly higher in the low mass sample than in the high mass sample.  On the other hand, as discussed in Section \ref{sec:compspecs}, uncertainties in the extinction law for young galaxies suggest that $E(B-V)$ for this sample may be overestimated, and independent measurements of extinction from {\it Spitzer} 24\micron\ imaging indicate that, on average, dust attenuation is larger in more massive galaxies \citep{rep+10}.  Even if this is the case, however, the magnitude of the effect does not appear to be sufficient.  \citet{pkr11} model the effects of dust as a function of the optical depth $\tau_{\rm dust}$, finding that flux is reduced by a factor of $\sim(1+\tau_{\rm dust})^{-1}$; in order to suppress the emission to the observed degree in the high mass and high dust samples, $\tau_{\rm dust}>3$ is required, while our values of $E(B-V)$ (which do not suffer from the systematic uncertainties affecting the lower mass sample) indicate $\tau_{\rm dust} \approx 1.5$--2 for these samples.  

\citet{pkr11} also consider a dust+ISM model. As discussed above, the effect of an ISM component is to increase both the absorption and total emission, while shifting the emission from the resonance lines to the fine structure transitions; if the ISM is also dusty, the emission is further reduced, particularly in the resonance lines. However, the dust+ISM model with $\tau_{\rm dust}=1$ still exhibits increased emission relative to the fiducial, dust-free model, indicating that the increase in both absorption and emission outweighs the effect of dust; this is
counter to our observations of weaker emission in more massive galaxies with more significant ISM. It may be that such a model with somewhat higher dust content would sufficiently suppress the emission without obscuring it entirely; more detailed modeling is needed, and some degree of fine tuning of the dust content would seem to be required to match the observations. For the moment we conclude that dust is an important factor but may not be the only explanation for the observed trend.

The remaining factor is the geometry of the absorbing and emitting regions.  Anisotropic outflows viewed from certain directions may produce observed absorption to emission ratios that deviate from the expected conservation of photons; in particular, a galaxy with a bi-conical outflow observed edge-on will show no absorption but strong emission. Thus observations of individual galaxies at a variety of inclinations could produce a trend similar to the observed $W_{\rm abs}$ vs.\ $W_{\rm em}$ relationship.  This is unlikely to be the explanation for the current sample, however, since our data points are for composite spectra of galaxies selected without regard to inclination or morphology.\footnote{Among local galaxies, selecting by dust content may introduce an inclination bias, since dustier galaxies are more likely to be edge-on. However, among the parent sample from which the galaxies observed here are drawn $E(B-V)$ does not increase with apparent inclination  (D.R. Law, private communication), and in any case such a bias would result in stronger emission and weaker absorption for the dustier subsample, the opposite of what is observed.}

\begin{figure*}[htbp]
\plotone{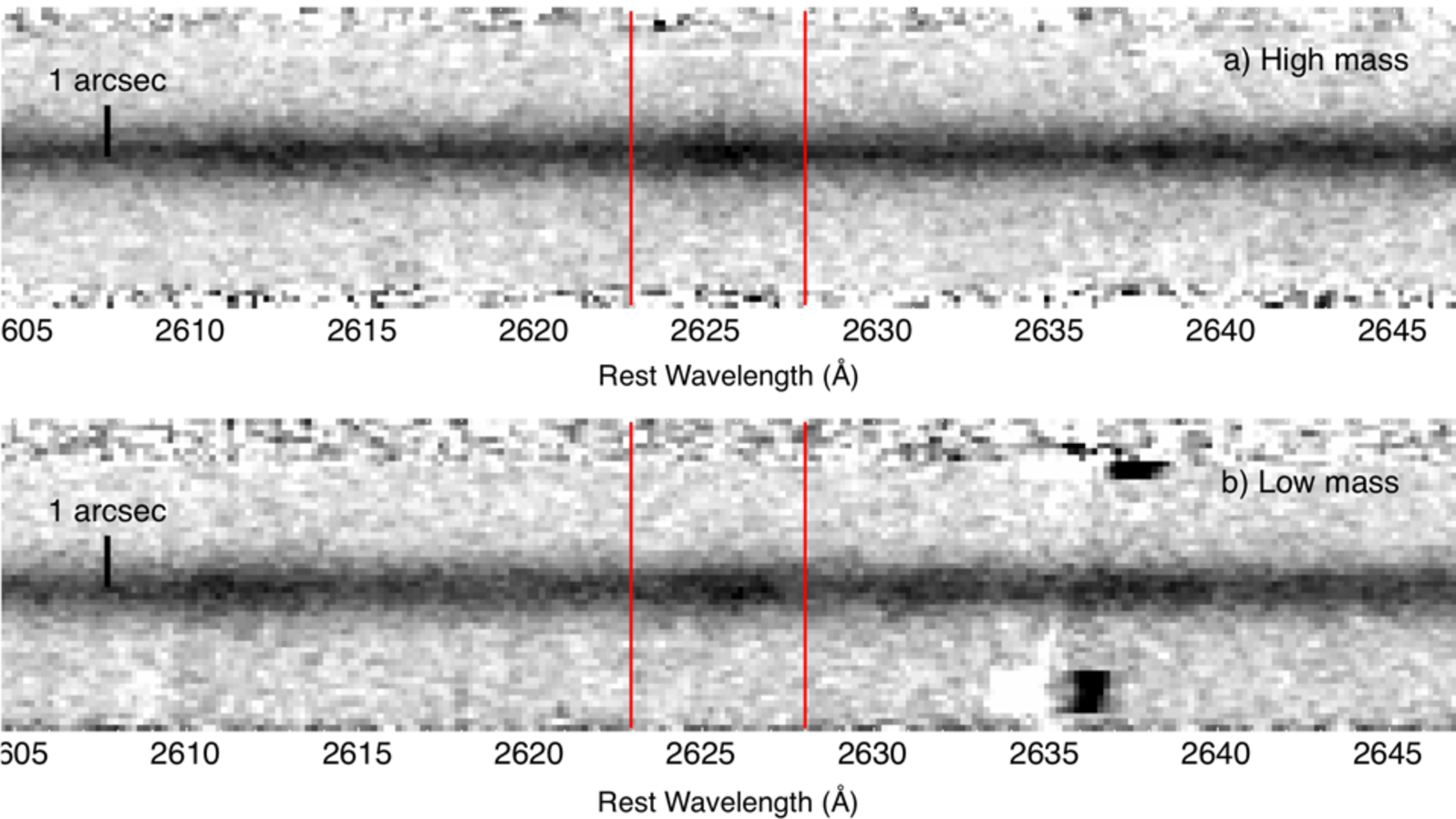}
\caption{Two-dimensional composite spectra of  \FeII* $\lambda2626$ emission in the high (top panel) and low (bottom panel) mass subsamples.  Spatial profiles are measured over the wavelength region indicated by the red vertical lines.}
\label{fig:highlowmass_2d}
\end{figure*}

\begin{figure*}[htbp]
\plotone{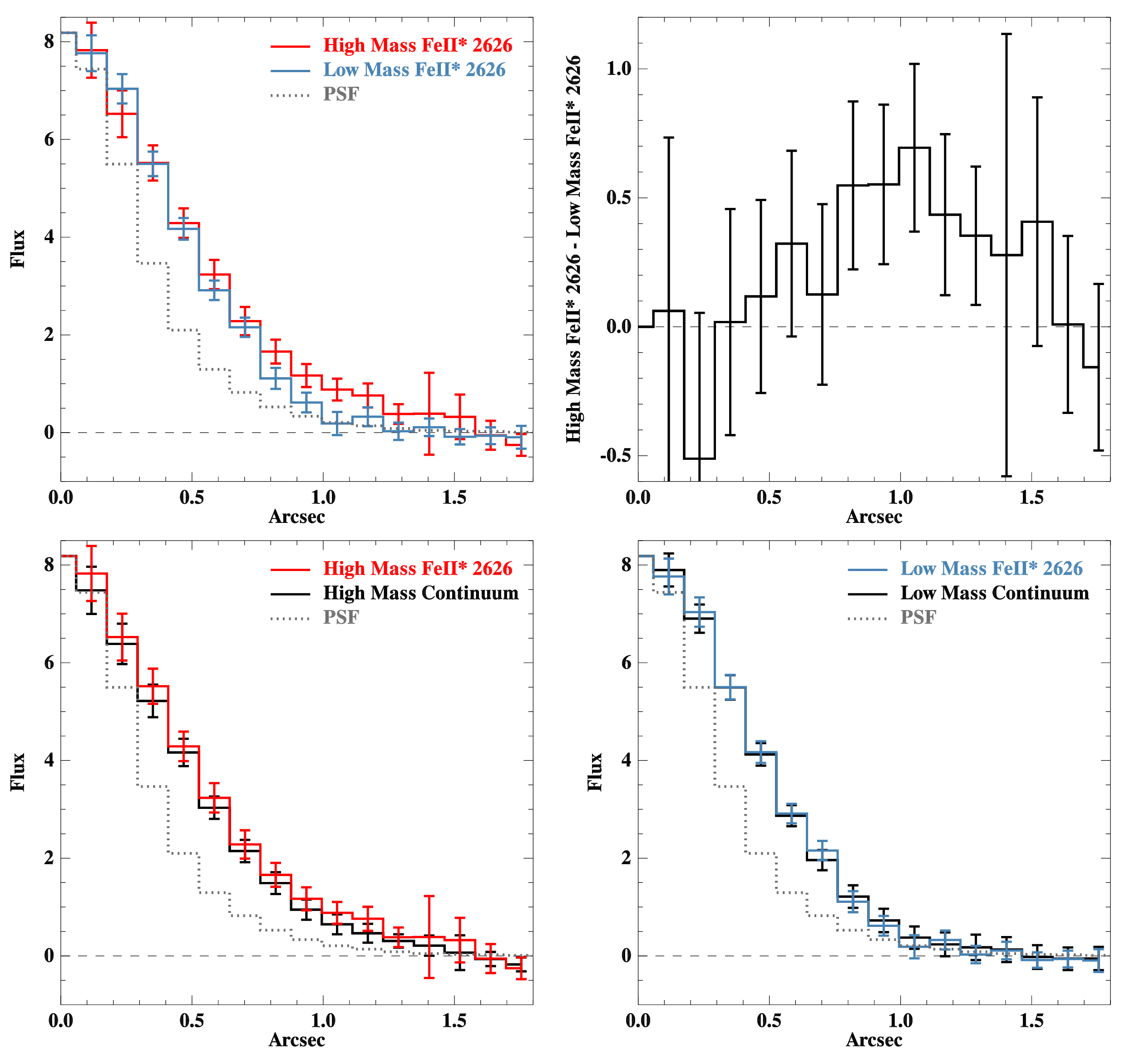}
\caption{{\it Top left}: Spatial profiles of the \FeII* $\lambda2626$ emission line in the two-dimensional composite spectra of the high mass (red) and low mass (blue) galaxy samples. {\it Top right}:  The excess flux in the high mass line emission relative to the low mass. {\it Bottom left:} Spatial profiles of the \FeII* $\lambda2626$ emission line (in red) and the continuum (in black) in the high mass subsample.   {\it Bottom right:} Spatial profiles of the \FeII* $\lambda2626$ emission line (in blue) and the continuum (in black) in the low mass subsample. }
\label{fig:spatialprofiles_mass}
\end{figure*}

The last issue we consider is the size of the \FeII*-emitting region relative to the area encompassed by the spectroscopic slit.  If fine structure emission occurs in outflowing gas at large distances from galaxies, the emitting region may be larger than the region subtended by the slit; the broader the extent of the emission, the higher the fraction that may fall outside the slit.  In this context, previous observations of low ionization absorption and fine structure emission in far-UV spectra at high redshift provide useful insights; the resonant \SiII\ absorption lines are coupled to non-resonant \SiII* emission in the same fashion as the \FeII\ lines studied in this work.  \citet{ssp+03} detected \SiII* fine structure emission accompanying \SiII\ absorption in the composite spectrum of Lyman break galaxies at $z\sim3$, finding that the emission lines were too strong to be produced in the \HII\ regions of the galaxies but much weaker than the \SiII\ absorption lines. Although they found no fully satisfactory explanation for the \SiII* lines (outflow models were deemed inadequate due to the narrowness of the \SiII* lines relative to \lya), they suggested that the weakness of the \SiII* lines could be due to the spectroscopic slit subtending only a small portion of the emitting region.  More recently, \citet{jse12}  compare the composite spectra of Lyman break galaxies at $z\sim4$ and $z\sim3$, and find that galaxies at $z\sim4$ have stronger \SiII* emission but weaker \SiII\ absorption than galaxies at $z\sim 3$, the same pattern we observe in the \FeII\ lines in low and high mass (and age, SFR, and dust) galaxies at $z\sim1.5$.  They interpret this pattern as evidence that \SiII\ photons are absorbed and re-emitted at larger distances at $z\sim3$ than at $z\sim4$.   

If slit losses are in part responsible for the weakness of the fine structure emission lines, significant absorption in the resonance lines must occur at large distances from the galaxies.  Recent results indicate that this is likely to be the case.  \citet{ses+10} use stacked spectra of paired foreground and background galaxies to measure the strength of low ionization absorption as a function of impact parameter, finding $W_{\rm abs} = 0.4$ \AA\ at a mean distance of 31 kpc in both the \SiII\ $\lambda1260$ and $\lambda1527$ lines.  The strength of  \SiII\ $\lambda1260$ remains the same at 63 kpc, and significant absorption in \lya\ is seen at $>100$ kpc.  \citet{sbs+11} also see low surface brightness \lya\ emission extending to $\sim 80$ kpc.  In a study of \MgII\ absorption systems and \OII\ emission at $0.4 < z < 1.3$, \citet{mwn+11} conclude that outflows are the primary source of \MgII\ absorption systems, and that outflowing gas can reach radii of $\sim50$ kpc.  These results indicate that significant absorption and emission are likely to take place beyond the radius subtended by the slit; the 1\arcsec\ and 1.2\arcsec\ slits used in our study subtend distances of 8--10 kpc for galaxies at $1<z<2$.

The models of \citet{pkr11} also assess the effect of slit loss.  The fraction of absorption and emission at large radius depends on the wind model considered; for most models, they find that $\sim1$\arcsec\ slits are sufficient to capture most of the emission in high redshift galaxies.  A notable exception is the radiation-pressure-driven wind model of \citet{mqt05}, for which a slit subtending $\sim$20 kpc is required to admit most of the scattered photons.

Given the detection of significant resonant absorption at large radius in similar galaxies at \ztwo, we suggest that, along with dust, increasing slit losses in massive galaxies may explain the observed trend of decreasing \FeII* $\lambda$2626 emission with increasing  \FeII\ $\lambda$2600 absorption.  In this scenario, the smaller ratio of emission to absorption strength observed in high mass (age, dust, SFR) galaxies is an indication that \FeII\ photons are absorbed and re-emitted at greater distances from the galaxy than in lower mass and less dusty objects.  This is consistent with the observation in Section \ref{sec:maxvels} that massive galaxies have higher wind velocities.   We also find support for this scenario in observations of a single low mass galaxy; \citet{eps+10} studied \SiII\ absorption and \SiII* emission in a compact, unreddened galaxy at $z=2.3$, finding that the emission and absorption strengths were roughly equal.  In this galaxy, very few of the \SiII* photons are lost to dust or fall outside the slit. 

A prediction of this slit-loss model is that  \FeII* $\lambda$2626 emission should be more extended in high mass galaxies than in low mass galaxies. We test this by constructing two-dimensional composite spectra of the high and low mass samples as described in Section \ref{sec:spatialextent}, again determining uncertainties from the standard deviation of the spatial profiles of 500 artificial 2D composite spectra created using bootstrap resampling.  The 2D spectra are shown in Figure \ref{fig:highlowmass_2d}, and the spatial profiles are shown in Figure \ref{fig:spatialprofiles_mass}. In the top two panels of Figure \ref{fig:spatialprofiles_mass} we show the \FeII* $\lambda$2626 spatial profiles at left and the difference between the profiles in the high and low mass samples at right. Although the uncertainties are significant, we find that at a radius of 1\arcsec\ the flux of the high mass sample is nearly 5 times that of the low mass sample; the difference in flux at 1\arcsec\  is $0.69\pm0.36$, for a significance of 1.9$\sigma$. We also observe excess emission to $\sim1.5$\arcsec, with lower significance.  

This difference between the emission line profiles in the high and low mass galaxies can be further clarified by comparisons of the \FeII* line emission and the continuum for the high and low mass samples. The relevant spatial profiles are shown in the lower two panels of Figure \ref{fig:spatialprofiles_mass}.  A comparison of the line and continuum emission in the high mass sample shows that the emission line flux is $\sim30$\% higher at a radius of 0.8--1.1 arcsec; this difference is not statistically significant, however, given the large uncertainties associated with the small high and low mass samples. A similar comparison of the line and continuum emission in the low mass sample shows that the two profiles are the same to within 5\%. Finally, we compare the continuum profiles of the high and low mass samples. Massive galaxies are observed to be larger than low mass galaxies (e.g.\ \citealt{lss+12}), and a comparison of the continuum profiles to the PSF in the lower two panels of Figure \ref{fig:spatialprofiles_mass} suggests that this is also the case in our sample. More quantitatively, at a radius of 0.8--1.1 arcsec the continuum emission from massive galaxies is $\sim40$\% higher than the continuum emission from low mass galaxies, but as with the line and continuum comparison above, this difference is not statistically significant.  

Thus the more extended line emission observed for high mass galaxies shown in the upper left panel of Figure \ref{fig:spatialprofiles_mass} appears to be due to a combination of extended line emission in massive galaxies and the larger sizes of massive galaxies; neither of these effects are statistically significant alone, but in combination they produce the moderately more significant difference between the \FeII* profiles in high and low mass galaxies.  In summary, the observation of more extended emission from the outflow in more massive galaxies supports the scenario in which slit losses are more significant for more massive galaxies.  A definitive test requires the detection of emission at much larger radii, but the weakness of the fine structure lines is likely to preclude this test for some time.  

A second important test of the model is a comparison of the strength of absorption as a function of impact parameter for both high and low mass galaxies, using pairs of foreground and background galaxies or galaxies with background quasars.  The scenario presented here predicts stronger low ionization metal line absorption at larger radii for more massive galaxies. Such results have already been found for galaxies associated with \MgII\ absorption systems at lower redshifts \citep{cwt+10}, and the coming years will see such studies extended to the redshifts probed by this sample and higher.

We have thus far focused discussion on the  \FeII\ $\lambda$2600 absorption and \FeII* $\lambda$2626 emission lines, as \FeII* $\lambda$2626 is the strongest of the \FeII* lines.  Although we do not detect all of the \FeII* lines in all of the composites, the other lines in the UV1 multiplet show the same trend of stronger emission and weaker absorption in low mass galaxies; we measure only upper limits on \FeII* $\lambda$2612 and \FeII* $\lambda$2632 in the high mass composites.  The UV2 multiplet presents a confusing picture, however. In this case we observe weaker absorption and stronger emission in the high mass and high dust samples, although galaxies with \MgII\ emission also show weaker absorption and stronger emission. The explanation for these mixed results is unclear.  In the UV3 multiplet we again observe weaker absorption in the low mass sample than in the high mass, but the  \FeII* $\lambda$2365 line is too weak to lead to useful conclusions regarding the emission.  As the UV2 observations partially conflict with our proposed model, further study of these lines is needed.

We have shown that the average velocity profile of \FeII* $\lambda$2626 emission, photoionization modeling, and the extent of the emission relative to the continuum support the scenario in which fine structure emission is generated by photon scattering in outflowing gas.  Greater extinction and/or a greater spatial extent of outflowing gas in massive galaxies is required to explain the observed decrease in \FeII* emission with increasing \FeII\ absorption.

\section{Summary and Conclusions}
\label{sec:summary}
We have presented measurements of rest-frame near-UV \FeII\ and \MgII\ absorption and emission in a sample of 96 star-forming galaxies at $1< z < 2$.  The presence of large-scale galactic outflows is indicated by the blueshift of interstellar absorption lines relative to the systemic velocity.  Our primary results are as follows.

\begin{enumerate}
\item{For 51 of the 96 galaxies, we obtain systemic redshifts from the centroids of strong \OII\ $\lambda\lambda$ 3726, 3729 emission, which falls at the red end of the spectral range covered.  These galaxies have an average redshift $\langle z \rangle = 1.4$.  Interstellar \FeII\ absorption lines in these galaxies show an average offset from the systemic velocity of $-85$ \kms, less than the average offset of $-166$ \kms\ measured for galaxies at \ztwo.   We use this offset of $-85$ \kms\ to estimate the systemic redshift of the remainder of the galaxies in the sample.  We also show that the average velocity offset decreases from $-200$ \kms\ at $z\sim2.4$ to $-110$ \kms\ at $z\sim2.1$.   These higher redshift samples have a slightly higher average mass than our $z\sim1.4$ sample, and have velocity offsets measured from far-UV interstellar absorption lines rather than the near-UV \FeII\ lines used in this study.  More careful accounting of these selection effects, and of the effects of the interstellar medium on the velocity measurements, is therefore required to determine whether the decrease in velocity offset with decreasing redshift represents a real trend of decreasing outflow velocity.} 

\item{Individual galaxies in the sample show more variety in the \MgII\ doublet than in the \FeII\ resonance lines.  \FeII\ is always seen in absorption, while \MgII\ ranges from strong emission to pure absorption.  Emission is more common in galaxies with bluer UV slopes, echoing far-UV observations of galaxies at $z\sim2$--4 in which \lya\ emission is stronger in bluer galaxies.  \MgII\ emission is also stronger in low mass galaxies, and fine structure \FeII* emission is stronger in galaxies with \MgII\ emission.} 

\item{We compare the velocities and equivalent widths of the \FeII\ and \MgII\ transitions in individual spectra, in order to assess the effect of emission filling on measurements of outflow velocity.  Underlying emission appears to shift the velocity centroid of the \FeII\ $\lambda 2383$ line to the blue in $\sim 1/3$ of the sample, while the \MgII\ $\lambda 2796$ line shows such a shift in $\sim 2/3$ of the sample. Studies of individual objects suggest that emission filling may be a stronger effect in low mass galaxies, but composite spectra show little difference in velocity centroids between the \FeII\ lines expected to be most and least affected by emission filling.  \MgII\ is much more clearly affected  by underlying emission.}

\item{The average maximum outflow velocity (the maximum extent of the blue wing of the interstellar absorption lines in the composite spectrum of all of the galaxies in the sample) is $v_{\rm max} = -730$ \kms, in good agreement with other studies of star-forming galaxies at $z\sim1.5$ -- 2.  $|v_{\rm max}|$ is 1.5 times larger in the high mass subset of galaxies than in the low mass subset, consistent with previous results indicating $v_{\rm wind} \sim M_{\star}^{0.17}$.  The velocity centroids $|v|$ are also blueshifted to higher velocities in the composite of high mass galaxies than in the low mass composite.  Unlike some previous studies, we see no trend between $v_{\rm max}$ and star formation rate. }

\item{We present the following evidence in support of the model in which \FeII* and \MgII\ emission is generated by photon scattering in outflowing gas:}
\begin{itemize}
\item{ The velocity profile of \FeII* $\lambda 2626$ emission in the composite spectrum of the full sample of galaxies shows a blueshifted tail very similar to that of the \FeII\ $\lambda2600$ absorption line, from which the emission is presumed to arise. }
\item{ Photoionization modeling indicates that very high values of the ionization parameter, $\log U > -2$, are required in order for the \FeII* emission to originate in \HII\ regions. The ionization parameters of the galaxies in the sample are unlikely to be this high, implying that \FeII* emission is produced by the re-emission of continuum photons absorbed in the \FeII\ resonance transitions in outflowing gas.}
\item{In contrast to \FeII* emission, photoionization modeling shows that the strength of \MgII\ emission is consistent with its production in \HII\ regions. We conclude that \MgII\ photons from \HII\ regions resonantly scatter in outflowing gas, producing the observed blueshifted absorption and redshifted emission profiles.}
\item{  We compare the spatial extent of \MgII\ and \FeII* emission relative to the continuum, and find that both emission lines are moderately more extended.  Higher S/N measurements are needed to more robustly confirm this trend and measure the true extent of the emission.}
\end{itemize}
\item{Using composite spectra, we observe a trend of {\it decreasing} \FeII* emission with increasing \FeII\ absorption.  This trend is the opposite of what one would naively expect, if the \FeII* photons are originally absorbed in the \FeII\ transitions.  We attribute this trend to a combination of increased extinction in more massive galaxies and more extended outflows in massive galaxies, leading to greater slit losses for \FeII* photons.  Two-dimensional composite spectra of high and low mass galaxies indicate that emission from the outflow is stronger at a radius of $\sim10$ kpc in high mass galaxies than in low mass galaxies.}
\end{enumerate}

Our results suggest that \MgII\ and \FeII* emission provide a valuable but so far little-utilized tool for the study of galactic outflows.  In particular, as telescope capabilities develop, studies of the spatial extent of these lines may allow important measurements of the physical extent of outflowing gas.  These lines are accessible from the ground over a wide range in redshift, from $z\sim 0.2$ to \ztwo, and complement the far-UV lines studied in galaxies at $z\sim 2$ and above.

\acknowledgements{We thank the referee for a thorough and constructive report which significantly improved the paper, Max Pettini for useful advice throughout the project, Ralf Kotulla for assistance with the construction of composite spectra, and Jay Gallagher, Claus Leitherer, Norm Murray, Alice Shapley and Christy Tremonti for helpful and interesting discussions.  Michael Cooper provided valuable assistance with the data reduction, and Chuck Steidel, Alice Shapley and Naveen Reddy made possible the survey from which the targets in this paper are drawn.  D.K.E. was supported in part by the Spitzer Fellowship Program of the National Aeronautics and Space Administration under Award No.\ NAS7-03001 and the California Institute of Technology. A.M.Q. was supported by the Marshall Scholarship and the National Science Foundation Graduate Research Fellowship.  C.L.M. and A.L.H. are supported by the National Science Foundation through grants AST-0808161 and AST-1109288 and by the David and Lucile Packard Foundation. The analysis pipeline used to reduce the DEIMOS data was developed at UC Berkeley with support from NSF grant AST-0071048.  We wish to extend thanks to those of Hawaiian ancestry on whose sacred mountain we are privileged to be guests.}

\bibliographystyle{apj}

\clearpage

\begin{turnpage}
\begin{deluxetable}{l c c c c c c c c c c c c}
\tablewidth{0pt}
\tabletypesize{\footnotesize}
\tablecaption{Measured Absorption Features in Composite Spectra\tablenotemark{a}\label{tab:abs}}
\tablehead{
\colhead{} &
\colhead{All} & 
\colhead{Low} & 
\colhead{High} & 
\colhead{Low} &
\colhead{High} &
\colhead{Low} &
\colhead{High} &
\colhead{Low} &
\colhead{High} &
\colhead{No MgII} &
\colhead{MgII} \\
\colhead{} &
\colhead{Galaxies} &
\colhead{Mass\tablenotemark{b}} & 
\colhead{Mass\tablenotemark{b}} & 
\colhead{E(B-V)\tablenotemark{c}} &
\colhead{E(B-V)\tablenotemark{c}} &
\colhead{SFR\tablenotemark{d}} &
\colhead{SFR\tablenotemark{d}} &
\colhead{Age\tablenotemark{e}} &
\colhead{Age\tablenotemark{e}} &
\colhead{Emission} &
\colhead{Emission} 
}
\startdata
$N$                &       96      & 24             & 24            &24             & 24            &24             & 24            &24             & 24            & 63            &33\\
\hline
$W_{{\rm FeII}\lambda2249}$  & $0.23 \pm 0.07$  & $<0.65$  & $0.31 \pm 0.08$  & $0.38 \pm 0.08$  & $<0.28$  & $<0.54$  & $<0.24$  & $<0.43$  & $<0.50$  & $0.40 \pm 0.09$  & $<0.18$  \\
$W_{{\rm FeII}\lambda2260}$  & $0.36 \pm 0.11$  & $<0.55$  & $<0.41$  & $0.47 \pm 0.13$  & $<0.56$  & $0.57 \pm 0.18$  & $<0.45$  & $<0.48$  & $<0.67$  & $0.51 \pm 0.12$  & $<0.21$  \\
$W_{{\rm FeII}\lambda2344}$  & $2.56 \pm 0.14$  & $2.00 \pm 0.12$  & $2.69 \pm 0.16$  & $2.52 \pm 0.18$  & $2.28 \pm 0.17$  & $2.27 \pm 0.20$  & $2.45 \pm 0.13$  & $2.03 \pm 0.14$  & $2.19 \pm 0.20$  & $2.60 \pm 0.18$  & $2.06 \pm 0.14$  \\
$W_{{\rm FeII}\lambda2374}$  & $1.77 \pm 0.12$  & $1.84 \pm 0.14$  & $1.53 \pm 0.11$  & $1.87 \pm 0.11$  & $1.47 \pm 0.10$  & $1.76 \pm 0.12$  & $1.47 \pm 0.08$  & $1.70 \pm 0.10$  & $1.70 \pm 0.18$  & $1.90 \pm 0.17$  & $1.25 \pm 0.09$  \\
$W_{{\rm FeII}\lambda2382}$  & $2.26 \pm 0.17$  & $1.98 \pm 0.20$  & $2.00 \pm 0.12$  & $1.95 \pm 0.11$  & $1.73 \pm 0.11$  & $1.79 \pm 0.12$  & $1.94 \pm 0.10$  & $1.73 \pm 0.13$  & $2.19 \pm 0.12$  & $2.45 \pm 0.23$  & $1.56 \pm 0.11$  \\
$W_{{\rm FeII}\lambda2586}$  & $2.31 \pm 0.11$  & $1.99 \pm 0.16$  & $2.80 \pm 0.15$  & $2.11 \pm 0.10$  & $2.76 \pm 0.16$  & $2.24 \pm 0.13$  & $2.48 \pm 0.11$  & $2.07 \pm 0.14$  & $2.70 \pm 0.19$  & $2.32 \pm 0.11$  & $2.14 \pm 0.13$  \\
$W_{{\rm FeII}\lambda2600}$  & $2.55 \pm 0.17$  & $2.58 \pm 0.24$  & $2.89 \pm 0.23$  & $2.34 \pm 0.10$  & $3.28 \pm 0.28$  & $2.63 \pm 0.20$  & $2.82 \pm 0.20$  & $2.94 \pm 0.27$  & $2.75 \pm 0.26$  & $2.64 \pm 0.17$  & $2.09 \pm 0.21$  \\
$W_{{\rm MgII}\lambda2796}$  & $1.60 \pm 0.12$  & $0.78 \pm 0.25$  & $3.14 \pm 0.15$  & $1.57 \pm 0.18$  & $1.95 \pm 0.25$  & $1.25 \pm 0.23$  & $1.70 \pm 0.17$  & $1.01 \pm 0.21$  & $2.44 \pm 0.33$  & $2.10 \pm 0.17$  & $1.17 \pm 0.08$  \\
$W_{{\rm MgII}\lambda2803}$  & $1.53 \pm 0.06$  & $0.79 \pm 0.25$  & $3.10 \pm 0.12$  & $1.63 \pm 0.09$  & $1.82 \pm 0.22$  & $1.40 \pm 0.13$  & $2.22 \pm 0.16$  & $1.13 \pm 0.26$  & $2.22 \pm 0.16$  & $1.95 \pm 0.11$  & $1.17 \pm 0.06$  \\
$W_{{\rm MgI}\lambda2852}$  & $0.53 \pm 0.09$  & $<0.71$  & $0.92 \pm 0.12$  & $0.49 \pm 0.14$  & $<0.64$  & $0.53 \pm 0.16$  & $0.65 \pm 0.19$  & $<0.60$  & $0.66 \pm 0.12$  & $0.49 \pm 0.16$  & $0.49 \pm 0.13$  \\
\hline
$v_{{\rm FeII}\lambda2249}$  & $ 24 \pm  56$  &  ...  & $ -2 \pm  40$  & $-21 \pm  40$  &  ...  &  ...  &  ...  &  ...  &  ...  & $ 82 \pm  47$  &  ...  \\
$v_{{\rm FeII}\lambda2260}$  & $-108 \pm  75$  &  ...  &  ...  & $-151 \pm  72$  &  ...  & $-136 \pm  93$  &  ...  &  ...  &  ...  & $-99 \pm  59$  &  ...  \\
$v_{{\rm FeII}\lambda2344}$  & $-71 \pm  33$  & $-31 \pm  17$  & $-97 \pm  33$  & $-73 \pm  42$  & $-52 \pm  31$  & $-123 \pm  41$  & $-55 \pm  23$  & $-55 \pm  26$  & $-105 \pm  42$  & $-101 \pm  38$  & $-32 \pm  31$  \\
$v_{{\rm FeII}\lambda2374}$  & $-120 \pm  27$  & $-39 \pm  26$  & $-127 \pm  23$  & $-149 \pm  25$  & $-82 \pm  16$  & $-128 \pm  22$  & $-95 \pm  13$  & $-62 \pm  18$  & $-140 \pm  45$  & $-119 \pm  36$  & $-93 \pm  22$  \\
$v_{{\rm FeII}\lambda2382}$  & $-118 \pm  46$  & $-37 \pm  40$  & $-114 \pm  20$  & $-176 \pm  25$  & $-46 \pm  20$  & $-189 \pm  25$  & $-108 \pm  16$  & $-62 \pm  24$  & $-103 \pm  19$  & $-112 \pm  52$  & $-143 \pm  26$  \\
$v_{{\rm FeII}\lambda2586}$  & $-91 \pm  19$  & $-53 \pm  24$  & $-120 \pm  27$  & $-67 \pm  12$  & $-116 \pm  24$  & $-48 \pm  18$  & $-148 \pm  15$  & $-106 \pm  21$  & $-59 \pm  32$  & $-102 \pm  17$  & $-69 \pm  21$  \\
$v_{{\rm FeII}\lambda2600}$  & $-189 \pm  39$  & $-131 \pm  43$  & $-176 \pm  42$  & $-124 \pm  11$  & $-173 \pm  47$  & $-139 \pm  29$  & $-163 \pm  36$  & $-171 \pm  48$  & $-173 \pm  45$  & $-156 \pm  33$  & $-171 \pm  49$  \\
$v_{{\rm MgII}\lambda2796}$  & $-281 \pm  23$  & $-230 \pm  55$  & $-255 \pm  15$  & $-306 \pm  32$  & $-217 \pm  43$  & $-270 \pm  33$  & $-245 \pm  26$  & $-238 \pm  47$  & $-254 \pm  43$  & $-189 \pm  24$  & $-337 \pm  17$  \\
$v_{{\rm MgII}\lambda2803}$  & $-203 \pm   7$  & $-162 \pm  33$  & $-193 \pm  11$  & $-188 \pm  10$  & $-231 \pm  31$  & $-204 \pm  17$  & $-190 \pm  20$  & $-164 \pm  24$  & $-190 \pm  19$  & $-149 \pm  11$  & $-260 \pm   8$  \\
$v_{{\rm MgI}\lambda2852}$  & $-130 \pm  35$  &  ...  & $-71 \pm  27$  & $-102 \pm  62$  &  ...  & $-134 \pm  46$  & $-159 \pm  55$  &  ...  & $-61 \pm  23$  & $-92 \pm  54$  & $-146 \pm  54$ 
\enddata
\tablenotetext{a}{Equivalent widths $W_0$ are given in \AA\ in the
  upper half of the table, and velocity centroids $v$ are given in
  \kms\ in the lower half of the table.}
\tablenotetext{b}{Minimum, maximum, mean and median stellar masses for
  the low mass composite: $1.3\times10^8$, $3.9\times10^9$,
  $1.9\times10^9$, $1.8\times10^9$ \msun\ respectively; minimum,
  maximum, mean and median stellar masses for the high mass composite:
  $1.1\times10^{10}$, $9.7\times10^{10}$, $2.4\times10^{10}$,
  $1.5\times10^{10}$ \msun\ respectively.}
\tablenotetext{c}{Minimum, maximum, mean and median values of $E(B-V)$
  for the low $E(B-V)$ composite: 0.07, 0.18, 0.14, 0.15 respectively; minimum,
  maximum, mean and median values of $E(B-V)$ for the high $E(B-V)$
  composite: 0.23, 0.46, 0.31, 0.29 respectively.}
\tablenotetext{d}{Minimum, maximum, mean and median SFRs for the low
  SFR composite: 5, 17, 10, 9 \msunyr\ respectively; minimum, maximum, mean and median SFRs for the high SFR composite: 31, 628, 128, 56 \msunyr\ respectively.}
\tablenotetext{e}{Minimum, maximum, mean and median ages for the low
  age composite: 7, 180, 55, 23 Myr respectively; minimum, maximum, mean and median ages for the high age composite: 571, 4750, 1260, 856 Myr respectively.}
\end{deluxetable}
\end{turnpage}

\clearpage

\begin{turnpage}
\setlength{\tabcolsep}{0.04in}\begin{deluxetable}{l c c c c c c c c c c c c}
\tablewidth{0pt}
\tabletypesize{\footnotesize}
\tablecaption{Measured Emission Features in Composite Spectra\tablenotemark{a}\label{tab:em}}
\tablehead{
\colhead{} &
\colhead{All} & 
\colhead{Low} & 
\colhead{High} & 
\colhead{Low} &
\colhead{High} &
\colhead{Low} &
\colhead{High} &
\colhead{Low} &
\colhead{High} &
\colhead{No MgII} &
\colhead{MgII} \\
\colhead{} &
\colhead{Galaxies} &
\colhead{Mass\tablenotemark{b}} & 
\colhead{Mass\tablenotemark{b}} & 
\colhead{E(B-V)\tablenotemark{c}} &
\colhead{E(B-V)\tablenotemark{c}} &
\colhead{SFR\tablenotemark{d}} &
\colhead{SFR\tablenotemark{d}} &
\colhead{Age\tablenotemark{e}} &
\colhead{Age\tablenotemark{e}} &
\colhead{Emission} &
\colhead{Emission} 
}
\startdata
$N$                &       96      & 24             & 24            &24             & 24            &24             & 24            &24             & 24            & 63            &33\\
\hline
$W_{{\rm CII]}\lambda2326}$  & $-0.74 \pm 0.07$  & $-1.18 \pm 0.29$  & $>-0.46$  & $-1.02 \pm 0.08$  & $>-0.96$  & $-0.97 \pm 0.16$  & $-0.90 \pm 0.18$  & $-0.83 \pm 0.27$  & $>-0.76$  & $-0.68 \pm 0.13$  & $-0.76 \pm 0.12$  \\
$W_{{\rm FeII}\lambda2365}$  & $>-0.22$  & $-0.23 \pm 0.07$  & $>-0.28$  & $-0.26 \pm 0.05$  & $>-0.32$  & $-0.35 \pm 0.07$  & $-0.30 \pm 0.10$  & $-0.23 \pm 0.07$  & $>-0.23$  & $>-0.32$  & $-0.31 \pm 0.07$  \\
$W_{{\rm FeII}\lambda2396}$  & $-0.40 \pm 0.05$  & $-0.27 \pm 0.07$  & $-0.70 \pm 0.09$  & $-0.45 \pm 0.06$  & $-0.61 \pm 0.12$  & $-0.78 \pm 0.08$  & $-0.57 \pm 0.10$  & $-0.35 \pm 0.08$  & $-0.47 \pm 0.10$  & $-0.32 \pm 0.07$  & $-0.61 \pm 0.08$  \\
$W_{{\rm FeII}\lambda2612}$  & $-0.50 \pm 0.05$  & $-0.57 \pm 0.10$  & $>-0.28$  & $-0.59 \pm 0.09$  & $-0.47 \pm 0.15$  & $-0.24 \pm 0.07$  & $>-0.51$  & $>-0.51$  & $>-0.26$  & $-0.33 \pm 0.07$  & $-0.84 \pm 0.10$  \\
$W_{{\rm FeII}\lambda2626}$  & $-0.90 \pm 0.05$  & $-0.94 \pm 0.10$  & $-0.73 \pm 0.08$  & $-1.01 \pm 0.08$  & $-0.69 \pm 0.10$  & $-1.17 \pm 0.09$  & $-0.89 \pm 0.10$  & $-0.76 \pm 0.10$  & $-0.75 \pm 0.08$  & $-0.78 \pm 0.07$  & $-1.18 \pm 0.09$  \\
$W_{{\rm FeII}\lambda2632}$  & $-0.19 \pm 0.05$  & $-0.39 \pm 0.07$  & $>-0.21$  & $-0.34 \pm 0.09$  & $>-0.15$  & $-0.40 \pm 0.12$  & $-0.24 \pm 0.08$  & $-0.22 \pm 0.07$  & $>-0.30$  & $>-0.18$  & $-0.37 \pm 0.12$  \\
$W_{{\rm MgII}\lambda2796}$  & $>-0.16$  & $-0.49 \pm 0.14$  & $>-0.06$  & $-0.44 \pm 0.07$  & $>-0.11$  & $>-0.35$  & $>-0.09$  & $-0.41 \pm 0.13$  & $>-0.10$  & $>-0.10$  & $-1.17 \pm 0.06$  \\
$W_{{\rm MgII}\lambda2803}$  & $>-0.17$  & $>-0.38$  & $>-0.06$  & $>-0.23$  & $>-0.13$  & $-0.25 \pm 0.07$  & $>-0.12$  & $>-0.22$  & $>-0.12$  & $>-0.13$  & $-0.71 \pm 0.07$  \\
\hline
$v_{{\rm CII]}\lambda2326}$  & $ 59 \pm  25$  & $  4 \pm  87$  & ...  & $ 95 \pm  24$  & ...  & $ 71 \pm  48$  & $ -3 \pm  59$  & $  6 \pm  96$  & ...  & $ 23 \pm  49$  & $ 45 \pm  41$  \\
$v_{{\rm FeII}\lambda2365}$  & ...  & $-23 \pm  29$  & ...  & $-84 \pm  23$  & ...  & $-63 \pm  27$  & $  6 \pm  48$  & $  2 \pm  28$  & ...  & ...  & $-54 \pm  35$  \\
$v_{{\rm FeII}\lambda2396}$  & $-33 \pm  20$  & $-42 \pm  28$  & $-17 \pm  33$  & $  5 \pm  20$  & $-119 \pm  41$  & $ 31 \pm  19$  & $-121 \pm  36$  & $-64 \pm  30$  & $  3 \pm  38$  & $-51 \pm  32$  & $-17 \pm  28$  \\
$v_{{\rm FeII}\lambda2612}$  & $-17 \pm  15$  & $ -1 \pm  24$  & ...  & $-51 \pm  25$  & $-34 \pm  61$  & $ 29 \pm  28$  & ...  & ...  & ...  & $-22 \pm  28$  & $-56 \pm  27$  \\
$v_{{\rm FeII}\lambda2626}$  & $-45 \pm   8$  & $-40 \pm  14$  & $ -3 \pm  18$  & $-29 \pm  14$  & $-45 \pm  20$  & $-56 \pm  10$  & $-31 \pm  30$  & $-52 \pm  20$  & $-40 \pm  15$  & $-47 \pm  12$  & $-44 \pm  17$  \\
$v_{{\rm FeII}\lambda2632}$  & $-15 \pm  28$  & $-10 \pm  16$  & ...  & $-34 \pm  37$  & ...  & $-21 \pm  30$  & $-20 \pm  48$  & $-18 \pm  26$  & ...  & ...  & $ 17 \pm  55$  \\
$v_{{\rm MgII}\lambda2796}$  & ...  & $  1 \pm  19$  & ...  & $-56 \pm  12$  & ...  & ...  & ...  & $  1 \pm  20$  & ...  & ...  & $-92 \pm   5$  \\
$v_{{\rm MgII}\lambda2803}$  & ...  & ...  & ...  & ...  & ...  & $ 19 \pm  20$  & ...  & ...  & ...  & ...  & $-43 \pm  17$  
\enddata
\tablenotetext{a}{Equivalent widths $W_0$ are given in \AA\ in the
  upper half of the table, and velocity centroids $v$ are given in
  \kms\ in the lower half of the table. Negative equivalent widths
  indicate emission.}
\tablenotetext{b}{Minimum, maximum, mean and median stellar masses for
  the low mass composite: $1.3\times10^8$, $3.9\times10^9$,
  $1.9\times10^9$, $1.8\times10^9$ \msun\ respectively; minimum,
  maximum, mean and median stellar masses for the high mass composite:
  $1.1\times10^{10}$, $9.7\times10^{10}$, $2.4\times10^{10}$,
  $1.5\times10^{10}$ \msun\ respectively.}
\tablenotetext{c}{Minimum, maximum, mean and median values of $E(B-V)$
  for the low $E(B-V)$ composite: 0.07, 0.18, 0.14, 0.15 respectively; minimum,
  maximum, mean and median values of $E(B-V)$ for the high $E(B-V)$
  composite: 0.23, 0.46, 0.31, 0.29 respectively.}
\tablenotetext{d}{Minimum, maximum, mean and median SFRs for the low
  SFR composite: 5, 17, 10, 9 \msunyr\ respectively; minimum, maximum, mean and median SFRs for the high SFR composite: 31, 628, 128, 56 \msunyr\ respectively.}
\tablenotetext{e}{Minimum, maximum, mean and median ages for the low
  age composite: 7, 180, 55, 23 Myr respectively; minimum, maximum, mean and median ages for the high age composite: 571, 4750, 1260, 856 Myr respectively.}  
\end{deluxetable}
\end{turnpage}

\end{document}